\documentclass[12pt]{article}
\usepackage{color}
\usepackage{amssymb,amsmath,comment}
\usepackage{graphicx}
\usepackage{braket}
\usepackage{epsfig}
\usepackage{here}
\graphicspath{{./Figures/}}
\newcommand{\ba}{\begin{align}}
\newcommand{\ea}{\end{align}}

\newcommand{\ov}{\overline}
\def\nn{\nonumber}

\def\bea{\begin{eqnarray}}
\def\eea{\end{eqnarray}}
 \makeatletter
\def\alt{\mathrel{\mathpalette\gl@align<}}
\def\agt{\mathrel{\mathpalette\gl@align>}}
\def\gl@align#1#2{\lower.6ex\vbox{\baselineskip\z@skip\lineskip\z@
\ialign{$\m@th#1\hfil##\hfil$\crcr#2\crcr\sim\crcr}}} \makeatother
\renewcommand{\thefootnote}{\fnsymbol{footnote}}
\begin{document}
\begin{flushright}
\end{flushright}
\vspace*{1.0cm}

\begin{center}
\baselineskip 20pt 
{\Large\bf 
Conditions for Suppressing
\\Dimension-five Proton Decay
\\in Renormalizable SUSY $SO(10)$ GUT
}
\vspace{1cm}

{\large 
Naoyuki Haba$^a$ \ and \ Toshifumi Yamada$^b$
} \vspace{.5cm}

{\baselineskip 20pt \it
$^a$ Department of Physics, Osaka Metropolitan University, Osaka 558-8585, Japan\\
$^b$ Department of Physics, Yokohama National University, Yokohama 240-8501, Japan
}

\vspace{.5cm}

\vspace{1.5cm} {\bf Abstract} \end{center}

The SUSY $SO(10)$ GUT is in severe tension with the experimental bounds on proton partial lifetimes
 because proton decay mediated by colored Higgsinos (dimension-five proton decay) is too rapid.
In this paper, we pursue the possibility that
 a texture of the Yukawa coupling matrices in a renormalizable SUSY $SO(10)$ GUT model suppresses dimension-five proton decay.
We focus on a general renormalizable SUSY $SO(10)$ GUT model 
 which contains ${\bf 10}+{\bf 126}+{\bf \ov{126}}+{\bf 120}$ representation fields
 and where the Yukawa coupling matrices of the {\bf 16} matter fields with 
 the ${\bf 10}$, ${\bf \ov{126}}$, ${\bf 120}$ fields, $Y_{10},Y_{126},Y_{120}$, 
 provide the quark and lepton Yukawa couplings and Majorana mass of the singlet neutrinos.
We find that if components in certain flavor bases, $(Y_{10})_{u_R d_R}$, $(Y_{126})_{u_R d_R}$,
 $(Y_{10})_{u_R s_R}$, $(Y_{126})_{u_R s_R}$, $(Y_{10})_{u_L d_L}$, $(Y_{126})_{u_L d_L}$, $(Y_{10})_{u_L s_L}$, $(Y_{126})_{u_L s_L}$, 
 $(Y_{10})_{u_L u_L}$, $(Y_{126})_{u_L u_L}$,
 are all on the order of the up quark Yukawa coupling, dimension-five proton decay can be suppressed
 while the Yukawa coupling matrices still reproduce the realistic quark and lepton masses and flavor mixings.
We numerically obtain specific Yukawa coupling matrices satisfying the above conditions,
 calculate proton partial lifetimes from them and evaluate how
 dimension-five proton decay is suppressed when these conditions are met.

\thispagestyle{empty}

\newpage
\renewcommand{\thefootnote}{\arabic{footnote}}
\setcounter{footnote}{0}
\baselineskip 18pt
\section{Introduction}

The $SO(10)$ grand unified theory (GUT)~\cite{Georgi:1974my,Fritzsch:1974nn} is a viable extension of the Standard Model (SM) 
 for its attractive features such as the embedding of the SM gauge groups into an anomaly-free group, the unification of one generation of the matter fields into a {\bf 16} representation field, 
 and the automatic realization of the seesaw mechanism~\cite{Gell-Mann:1979vob,Yanagida:1979as,Yanagida:1979gs,seesaw4} that naturally explains the tiny neutrino mass.
The supersymmetric (SUSY) $SO(10)$ GUT can further alleviate the gauge hierarchy problem,
 and achieve the gauge coupling unification without intermediate scale.
A drawback of the SUSY $SO(10)$ GUT is that proton decay mediated by colored Higgsinos (dimension-five proton decay)~\cite{Weinberg:1981wj,Sakai:1981pk}
 is too rapid to be consistent with the current experimental bounds on proton partial lifetimes.
In particular, since the unification of the top and bottom quark Yukawa couplings implies $\tan\beta\sim50$, the contribution of $E^cU^cU^cD^c$ operators~\cite{Goto:1998qg} to the $p\to K^+ \bar{\nu}_\tau$ decay is enhanced, and since simultaneous cancellations of $E^cU^cU^cD^c$ and $QQQL$ operators' contributions to $p\to K^+ \bar{\nu}_\tau$, and $QQQL$ operators' contributions to $p\to K^+ \bar{\nu}_\mu$ are difficult to realize, the SUSY $SO(10)$ GUT is in severe tension with the experimental bound on the $p\to K^+ \bar{\nu}$ partial lifetime~\cite{Super-Kamiokande:2014otb}.

However, there is a possibility that a texture of the Yukawa coupling matrices
 suppresses the troublesome dimension-five proton decay, because the decay amplitudes are proportional to bi-products of Yukawa couplings.
In this paper, we pursue the above possibility and find conditions for a texture of the Yukawa coupling matrices suppressing dimension-five proton decay.
We further obtain specific Yukawa coupling matrices that satisfy the conditions.
We focus on a general renormalizable $SO(10)$ GUT model which contains ${\bf 10}+{\bf 126}+{\bf \ov{126}}+{\bf 120}$ representation fields from which the Higgs fields of the minimal SUSY SM (MSSM) originate
 (a broader class of renormalizable SUSY $SO(10)$ GUT models have been studied in Refs.~\cite{Matsuda:2000zp}-\cite{Haba:2020ebr}).
In the model, the {\bf 16} matter fields have Yukawa couplings with the ${\bf 10},{\bf \ov{126}},{\bf 120}$ fields,
 which provide the quark and lepton Yukawa couplings and Majorana mass of the singlet neutrinos.
Note that the Yukawa couplings with the ${\bf 10},{\bf \ov{126}},{\bf 120}$ fields are most general, 
 since ${\bf 16}\times {\bf 16}={\bf 10}+{\bf 126}+{\bf 120}$ and they are the only allowed renormalizable couplings involving a pair of {\bf 16} matter fields.

In the main body of the paper, we identify those components of the Yukawa coupling matrices that are involved in dimension-five proton decay and that can be on the order of the up quark Yukawa coupling without contradicting that they give the realistic quark and lepton Yukawa couplings.
Here the up quark Yukawa coupling is considered as the smallest scale of the components of the Yukawa coupling matrices
 because it is a specially small Yukawa coupling in the SUSY $SO(10)$ GUT where $\tan\beta\sim50$.
That the components identified above be on the order of the up quark Yukawa coupling, is the desired
 conditions for a texture suppressing dimension-five proton decay.
Next, we obtain specific Yukawa coupling matrices satisfying these conditions,
 by fitting the experimental data of quark and lepton masses and flavor mixings
 with the Yukawa coupling matrices of the ${\bf 10},{\bf \ov{126}},{\bf 120}$ fields
 under the constraint that the components identified above be on the order of the up quark Yukawa coupling.
Then we calculate proton partial lifetimes from these Yukawa coupling matrices, and compare them with those calculated from Yukawa coupling matrices that do not necessarily satisfy the conditions. Thereby we evaluate how dimension-five proton decay is suppressed owing to the conditions.

Previously, suppression of dimension-five proton decay by a texture of the Yukawa coupling matrices in the SUSY $SO(10)$ GUT
 has been studied in Refs.~\cite{Dutta:2004zh,Dutta:2005ni,Haba:2020ebr}.
Those papers deal with the case when the active neutrino mass is dominated by the contribution of the Type-2 seesaw mechanism
 coming from a tiny VEV of the $SU(2)_L$-triplet component of the ${\bf \ov{126}}$ field.
However, the dominance of the Type-2 seesaw contribution is not a general situation, 
 since it requires a fine-tuning of a mass term, coupling constants
 and VEVs of GUT-breaking fields~\cite{Haba:2020ebr,Goh:2004fy} so as not to spoil the gauge coupling unification.
Thus, the present paper considers the case when the active neutrino mass is generated solely by the Type-1 seesaw mechanism,
 with singlet neutrinos coming from the {\bf 16} matter fields and their Majorana mass from the GUT-breaking VEV of the ${\bf \ov{126}}$ field.

This paper is organized as follows:
In Section~\ref{section-model}, we review the general renormalizable SUSY $SO(10)$ GUT model containing ${\bf 10}+{\bf 126}+{\bf \ov{126}}+{\bf 120}$ fields,
 and write the formulas for partial widths of dimension-five proton decay.
In Section~\ref{section-identification}, we derive conditions for a texture of the Yukawa coupling matrices suppressing dimension-five proton decay.
It will turn out that not only the texture of the Yukawa coupling matrices, but also a certain texture of the colored Higgs mass matrix is needed to suppress dimension-five proton decay. The latter texture is studied in Section~\ref{section-coloredhiggs}.
In Section~\ref{section-fitting}, we numerically obtain specific Yukawa coupling matrices satisfying the conditions found in Section~\ref{section-identification}.
We further calculate proton partial lifetimes from them and evaluate how dimension-five proton decay is suppressed when these conditions are met.
Section~\ref{summary} summarizes the paper.
\\

\section{General Renormalizable SUSY $SO(10)$ GUT Model}
\label{section-model}

\subsection{Model Description}

We consider a SUSY $SO(10)$ GUT model that contains 
 single {\bf 10}, single pair of ${\bf 126}+{\bf \ov{126}}$, single {\bf 120} fields, denoted by $H,\Delta+\ov{\Delta},\Sigma$, respectively.
The matter fields of MSSM and a singlet neutrino of each generation are unified into a ${\bf 16}$ representation field, denoted by ${\bf 16}^i$ with $i$ being the flavor index.
The Yukawa couplings are given by
\bea
W\ =\ (\tilde{Y}_{10})_{ij}\,{\bf 16}^i H{\bf 16}^j+(\tilde{Y}_{126})_{ij}\,{\bf 16}^i\overline{\Delta}{\bf 16}^j+(\tilde{Y}_{120})_{ij}\,{\bf 16}^i\Sigma{\bf 16}^j,
\label{gutyukawa}
\eea
 where $\tilde{Y}_{10}$, $\tilde{Y}_{126}$, $\tilde{Y}_{120}$ are Yukawa coupling matrices in the flavor space,
 and $\tilde{Y}_{10}$, $\tilde{Y}_{126}$ are complex symmetric and $\tilde{Y}_{120}$ is complex antisymmetric.
The quark and lepton Yukawa couplings are assumed to arise solely from Eq.~(\ref{gutyukawa}).

Additionally, we introduce single {\bf 210} and single {\bf 45} fields, denoted by $\Phi,A$, respectively.
The $\Phi,A$ develop vacuum expectation values (VEVs) to break $SU(5)$ subgroup of the $SO(10)$ while $\Delta+\ov{\Delta}$ develop VEVs to break $U(1)$ subgroup.

When the $SO(10)$ is broken into the SM gauge groups $SU(3)_C\times SU(2)_L\times U(1)_Y$, the $({\bf1},{\bf2},\pm\frac{1}{2})$ components of $H,\Delta,\ov{\Delta},\Sigma,\Phi$ yield the Higgs fields of MSSM.
Accordingly, $\tilde{Y}_{10},\tilde{Y}_{126},\tilde{Y}_{120}$ give 
 the up-type quark, down-type quark, charged lepton and neutrino Dirac Yukawa coupling matrices, $Y_u,Y_d,Y_e,Y_D$, as
\bea
W_{\rm eff}\ =\ (Y_u)_{ij}\,Q_i H_u U_j^c+(Y_d)_{ij}\,Q_i H_d D_j^c+(Y_e)_{ij}\,L_i H_d E_j^c+(Y_D)_{ij}\,L_i H_u N_j^c,
\label{mssmyukawa}
\eea
 where the Yukawa coupling matrices satisfy at a GUT breaking scale $\mu_{\rm GUT}$ the following relations:
\bea
&&Y_u\ =\ Y_{10}+r_2\,Y_{126}+r_3\,Y_{120},
\label{yu}\\
&&Y_d\ =\ r_1\left(Y_{10}+Y_{126}+Y_{120}\right),
\label{yd}\\
&&Y_e\ =\ r_1\left(Y_{10}-3Y_{126}+r_e\,Y_{120}\right),
\label{ye}\\
&&Y_D\ =\ Y_{10}-3r_2\,Y_{126}+r_D\,Y_{120},
\label{ydirac}
\eea
 where $Y_{10}\propto\tilde{Y}_{10}$, $Y_{126}\propto\tilde{Y}_{126}$, $Y_{120}\propto\tilde{Y}_{120}$,
 and $r_1,r_2,r_3,r_e,r_D$ are complex numbers determined from the mass matrix of the $({\bf1},{\bf2},\pm\frac{1}{2})$ components.
Hereafter we perform a phase redefinition of fields to make $r_1$ real positive.

The GUT-breaking VEV of $\ov{\Delta}$, denoted by $\ov{v}_R$, provides the singlet neutrinos with Majorana mass as
\bea
W_{\rm Majorana} \ =\ \frac{1}{2}(M_N)_{ij} \,N_i^c\,N_j^c, \ \ \ \ \ M_N \propto \tilde{Y}_{126}\,\ov{v}_R.
\label{majorana}
\eea
Integrating out $N_i^c$'s, we get the Weinberg operator
\bea
W_{\rm eff} \ =\ \frac{1}{2}(C_\nu)_{ij}\, L_i H_u L_j H_u,
\label{weinberg}
\eea
 where the Wilson coefficient $C_\nu$ satisfies at the scale of the singular values of $M_N$,
\bea
C_\nu \ =\ -Y_DM_N^{-1}Y_D^T.
\label{cnu}
\eea
Eq.~(\ref{weinberg}) gives rise to the Type-1 seesaw contribution to the active neutrino mass.
In this paper, we assume that the VEV of the $({\bf 1},{\bf 3},1)$ component of $\ov{\Delta}$ is so small that the Type-2 seesaw contribution is negligible compared to the Type-1 seesaw one.
\\

\subsection{Dimension-five Proton Decay}

The $({\bf 3},{\bf1},-\frac{1}{3})+({\bf \overline{3}}, {\bf1},\frac{1}{3})$ components of $H$, $\Delta$, $\ov{\Delta}$, $\Sigma$, $\Phi$,
 which we call colored Higgs fields and denote by $H_C^A,\ov{H}_C^B$ ($A,B$ are labels), induce dimension-five operators responsible for proton decay.
After the GUT breaking, the colored Higgs fields have GUT-scale mass ${\cal M}_{H_C}$ and Yukawa couplings with matter fields as
\begin{align}
W_{\rm colored \, Higgs}
&=\sum_{A,B} \ \ov{H}_C^A ({\cal M}_{H_C})_{AB} H_C^B
\nn\\
&+\sum_{A}\ \frac{1}{2}(Y_L^A)_{ij}Q_i H_C^A Q_j +  (\overline{Y}^A_L)_{ij}Q_i \overline{H}^A_C L_j + (Y_R^A)_{ij}E^c_i H_C^A U^c_j +  (\overline{Y}_R^A)_{ij}U^c_i \overline{H}_C^A D^c_j.
\label{coloredHiggs}
\end{align}
Here $Y_L^A,\ov{Y}_L^A,\,Y_R^A,\,\ov{Y}^A_R \propto Y_{10}$ when $H_C^A,\overline{H}^A_C$ are components of $H$,
 and they are proportional to $Y_{126}$ when $H_C^A,\overline{H}^A_C$ are components of $\ov{\Delta}$.
When $H_C^A,\overline{H}^A_C$ are components of $\Sigma$, Yukawa couplings $\ov{Y}_L^A,\,Y_R^A,\,\ov{Y}^A_R$ are proportional to $Y_{120}$ while $Y_L^A$ vanishes because $Y_L^A$ must be symmetric.
In the other cases, all the Yukawa couplings vanish.
By integrating out the colored Higgs fields, we obtain dimension-five operators responsible for proton decay
\bea
W_5 \ = \ -\frac{1}{2}C_{5L}^{ijkl}\,(Q_k Q_{l})(Q_i L_{j}) - C_{5R}^{ijkl}\,E^c_k U^c_{l} U^c_i D^c_{j},
\label{w5}
\eea
 where isospin indices are summed in each bracket in the first term.
The Wilson coefficients satisfy at the scale of the singular values of ${\cal M}_{H_C}$,
\begin{align}
C_{5L}^{ijkl} \ &=\ \sum_{A,B}({\cal M}_{H_C}^{-1})_{AB}
\left\{(Y_L^A)_{kl}(\ov{Y}^B_L)_{ij}
-\frac{1}{2}(Y_L^A)_{li}(\ov{Y}^B_L)_{kj}-\frac{1}{2}(Y_L^A)_{ik}(\ov{Y}^B_L)_{lj}\right\},
\label{c5lgeneral}\\
C_{5R}^{ijkl} \ &=\ \sum_{A,B}({\cal M}_{H_C}^{-1})_{AB}
\left\{(Y_R^A)_{kl}(\ov{Y}_R^B)_{ij}-(Y_R^A)_{ki}(\ov{Y}_R^B)_{lj}\right\}.
\label{c5rgeneral}
\end{align}

We write the partial widths of the $p\to K^+ \bar{\nu}_\tau$, $p\to K^+ \bar{\nu}_\mu$, $p\to K^+ \bar{\nu}_e$, $p\to K^0 \mu^+$, $p\to K^0 e^+$ decays induced by dimension-five operators (other decay modes will be commented on in the last paragraph of Section~\ref{section-identification}).
The partial widths read, for $\beta=e,\mu$,~\cite{Nath:2006ut}
\begin{align}
\Gamma(p\to K^+\bar{\nu}_\tau)
 = &\frac{m_N}{64\pi}\left(1-\frac{m_K^2}{m_N^2}\right)^2
\left\vert \beta_H(\mu_{\rm had})\frac{1}{f_\pi}\left\{
\left(1+\frac{D}{3}+F\right)C_{LL}^{s\tau \,ud}(\mu_{\rm had})+\frac{2D}{3}C_{LL}^{d\tau \,us}(\mu_{\rm had})\right\}
\right.
\nn\\
&\left. \ \ \ \ \ \ \ \ \ \ \ \ +\alpha_H(\mu_{\rm had})\frac{1}{f_\pi}\left\{
\left(1+\frac{D}{3}+F\right)C_{RL}^{ud \,\tau s}(\mu_{\rm had})+\frac{2D}{3}C_{RL}^{us \,\tau d}(\mu_{\rm had})
\right\}\right\vert^2,
\label{ptoknutau}
\\
\Gamma(p\to K^+\bar{\nu}_{\beta})
= &\frac{m_N}{64\pi}\left(1-\frac{m_K^2}{m_N^2}\right)^2
\left\vert \beta_H(\mu_{\rm had})\frac{1}{f_\pi}\left\{
\left(1+\frac{D}{3}+F\right)C_{LL}^{s\beta \,ud}(\mu_{\rm had})+\frac{2D}{3}C_{LL}^{d\beta \,us}(\mu_{\rm had})\right\}
\right\vert^2,
\label{ptoknumu}
\\
\Gamma(p\to K^0\beta^+)
= &\frac{m_N}{64\pi}\left(1-\frac{m_K^2}{m_N^2}\right)^2
\left\vert \beta_H(\mu_{\rm had})\frac{1}{f_\pi}
\left(1-D+F\right)\ov{C}_{LL}^{u\beta \,us}(\mu_{\rm had})
\right\vert^2,
\label{ptok0mu}
\end{align}
 where the Wilson coefficients of dimension-six operators $C_{RL},C_{LL},\ov{C}_{LL}$ satisfy\footnote{
  By writing $C_{5L}^{s\alpha\, ud}$, we mean that $Q_i$ is in the flavor basis where the down-type quark Yukawa coupling is
  diagonalized and that the down-type quark component of $Q_i$ is exactly $s$ quark.
  Likewise, $Q_k$ is in the flavor basis where the up-type quark Yukawa coupling is
  diagonalized and its up-type component is exactly $u$ quark, 
  and $Q_l$ is in the flavor basis where the down-type quark Yukawa coupling is
  diagonalized and its down-type quark component is exactly $d$ quark.
  The same rule applies to other Wilson coefficients.
  }, for $\alpha=e,\mu,\tau$,
\begin{align}
&C_{RL}^{ud\,\tau s}(\mu_{\rm had}) = A_{RL}(\mu_{\rm had},\mu_{\rm SUSY})
\frac{\mu_H}{m_{\tilde{t}_R}^2}
\frac{1}{16\pi^2}{\cal F}'(V^{\rm ckm}_{ts})^*\, y_t y_\tau
\,C_{5R}^{ud\tau t}\vert_{\mu=\mu_{\rm SUSY}},
\label{crl1}
\\
&C_{RL}^{us\,\tau d}(\mu_{\rm had}) = A_{RL}(\mu_{\rm had},\mu_{\rm SUSY})
\frac{\mu_H}{m_{\tilde{t}_R}^2}
\frac{1}{16\pi^2}{\cal F}'(V^{\rm ckm}_{td})^*\, y_t y_\tau
\,C_{5R}^{us\tau t}\vert_{\mu=\mu_{\rm SUSY}}
\label{crl2}
\\
&C_{LL}^{s\alpha\,ud}(\mu_{\rm had}) = A_{LL}(\mu_{\rm had},\mu_{\rm SUSY})
\frac{M_{\widetilde{W}}}{m_{\tilde{q}}^2}
\frac{1}{16\pi^2}{\cal F}\,
g_2^2\left(C_{5L}^{s\alpha\,ud}-C_{5L}^{u\alpha\, ds}\right)\vert_{\mu=\mu_{\rm SUSY}},
\label{cll1}
\\
&C_{LL}^{d\alpha\,us}(\mu_{\rm had}) = A_{LL}(\mu_{\rm had},\mu_{\rm SUSY})
\frac{M_{\widetilde{W}}}{m_{\tilde{q}}^2}
\frac{1}{16\pi^2}{\cal F}\,
g_2^2\left(C_{5L}^{d\alpha\, us}-C_{5L}^{u\alpha\, ds}\right)\vert_{\mu=\mu_{\rm SUSY}},
\label{cll2}
\\
&\ov{C}_{LL}^{u\beta\,us}(\mu_{\rm had}) = A_{LL}(\mu_{\rm had},\mu_{\rm SUSY})
\frac{M_{\widetilde{W}}}{m_{\tilde{q}}^2}
\frac{1}{16\pi^2}{\cal F}\,
g_2^2\left(-C_{5L}^{u\beta\, us}+C_{5L}^{s\beta\, uu}\right)\vert_{\mu=\mu_{\rm SUSY}},
\label{cll3}
\end{align}
 and the Wilson coefficients of dimension-five operators satisfy
\begin{align}
&C_{5R}^{ud\tau t}(\mu_{\rm SUSY}) = 
A_R^{\tau t}(\mu_{\rm SUSY},\mu_{\rm GUT})\sum_{A,B}({\cal M}_{H_C}^{-1})_{AB}\left.
\left\{(Y_R^A)_{\tau t}(\ov{Y}^B_R)_{ud}-(Y_R^A)_{\tau u}(\ov{Y}^B_R)_{td}\right\}\right|_{\mu=\mu_{\rm GUT}},
\label{c5r1}
\\
&C_{5R}^{us\tau t}(\mu_{\rm SUSY}) = 
A_R^{\tau t}(\mu_{\rm SUSY},\mu_{\rm GUT})\sum_{A,B}({\cal M}_{H_C}^{-1})_{AB}\left.
\left\{(Y_R^A)_{\tau t}(\ov{Y}^B_R)_{us}-(Y_R^A)_{\tau u}(\ov{Y}_R^B)_{ts}\right\}\right|_{\mu=\mu_{\rm GUT}},
\label{c5r2}
\\
&C_{5L}^{s\alpha\, ud}(\mu_{\rm SUSY})-C_{5L}^{u\alpha\, ds}(\mu_{\rm SUSY})
\nn\\
&=A_L^\alpha(\mu_{\rm SUSY},\mu_{\rm GUT})\sum_{A,B}({\cal M}_{H_C}^{-1})_{AB}\,\left.
\frac{3}{2}\left\{(Y_L^A)_{ud}(\ov{Y}_L^B)_{s\alpha}-(Y_L^A)_{ds}(\ov{Y}_L^B)_{u\alpha}\right\}\right|_{\mu=\mu_{\rm GUT}},
\label{c5l1}\\
&C_{5L}^{d\alpha\, us}(\mu_{\rm SUSY})-C_{5L}^{u\alpha\, ds}(\mu_{\rm SUSY})
\nn\\
&=A_L^\alpha(\mu_{\rm SUSY},\mu_{\rm GUT})\sum_{A,B}({\cal M}_{H_C}^{-1})_{AB}\,\left.
\frac{3}{2}\left\{(Y_L^A)_{us}(\ov{Y}_L^B)_{d\alpha}-(Y_L^A)_{ds}(\ov{Y}_L^B)_{u\alpha}\right\}\right|_{\mu=\mu_{\rm GUT}},
\label{c5l2}
\\
&C_{5L}^{u\beta \,us}(\mu_{\rm SUSY})-C_{5L}^{s\beta \,uu}(\mu_{\rm SUSY})
\nn\\ 
&=A_L^\beta(\mu_{\rm SUSY},\mu_{\rm GUT})\sum_{A,B}({\cal M}_{H_C}^{-1})_{AB}\,\left.
\frac{3}{2}
\left\{(Y_L^A)_{us}(\ov{Y}_L^B)_{u\beta}-(Y_L^A)_{uu}(\ov{Y}_L^B)_{s\beta}\right\}\right|_{\mu=\mu_{\rm GUT}},
\label{c5l3}
\end{align}
 where $\mu_{\rm had}$ denotes a hadronic scale, $\mu_{\rm SUSY}$ a soft SUSY breaking scale and $\mu_{\rm GUT}$ a GUT- breaking scale.
Here $\alpha_H,\beta_H$ denote the hadronic matrix elements, $D,F$ are parameters of the baryon chiral Lagrangian,
 and $C_{LL},\ov{C}_{LL},C_{RL}$ are the Wilson coefficients of the effective Lagrangian where the SUSY particles are integrated out,
 $-{\cal L}_6 = C_{LL}^{ijkl}(\psi_{u_Lk}\psi_{d_Ll})(\psi_{d_Li}\psi_{\nu_Lj})+\ov{C}_{LL}^{ijkl}(\psi_{u_Lk}\psi_{d_Ll})(\psi_{u_Li}\psi_{e_Lj})+C_{RL}^{ijkl}(\psi_{\nu_Lk}\psi_{d_Ll})(\psi_{u_R^ci}\psi_{d_R^cj})$  
 ($\psi$ denote SM Weyl spinors, and spinor indices are summed in each bracket).
In Eqs.~(\ref{crl1})-(\ref{cll3}), $y_t,y_\tau,g_2$ denote the top quark Yukawa, tau lepton Yukawa and weak gauge couplings in MSSM, respectively, and $V^{\rm ckm}_{ij}$ denotes $(i,j)$-component of CKM matrix.
${\cal F}',{\cal F}$ are loop functions defined as ${\cal F}' =\frac{1}{x-y}(\frac{x}{1-x}\log x - \frac{y}{1-y}\log y)$
 and ${\cal F} =\frac{1}{z-w}(\frac{z}{1-z}\log z - \frac{w}{1-w}\log w) + \frac{1}{z-1}(\frac{z}{1-z}\log z+1)$,
 where $x=|\mu_H|^2/m_{\tilde{t}_R}^2$, $y=m_{\tilde{\tau}_R}^2/m_{\tilde{t}_R}^2$, $z=|M_{\widetilde{W}}|^2/m_{\tilde{q}}^2$, $w=m_{\tilde{\ell}_\alpha}^2/m_{\tilde{q}}^2$,
 and $\mu_H,m_{\tilde{t}_R},m_{\tilde{\tau}_R},M_{\widetilde{W}},m_{\tilde{\ell}_\alpha},m_{\tilde{q}}$ denote the pole masses of Higgsinos, isospin-singlet top squark, isospin-singlet tau slepton,
 Winos, isospin-doublet slepton of flavor $\alpha$, and 1st and 2nd generation isospin-doublet squarks, respectively.
$A_{LL},A_{RL}$ account for corrections from renormalization group (RG) evolutions in SM from scale $\mu_{\rm SUSY}$ to $\mu_{\rm had}$.
Here RG corrections involving SM Yukawa couplings other than the top quark one are neglected and thus $A_{LL},A_{RL}$ are flavor-universal.
$A_R^{\tau t},A_L^\alpha$ account for corrections from RG evolutions in MSSM from scale $\mu_{\rm GUT}$ to $\mu_{\rm SUSY}$.

We rewrite the flavor-dependent part of Eqs.~(\ref{c5r1})-(\ref{c5l3}) with the GUT Yukawa coupling matrices $Y_{10},Y_{126},Y_{120}$ as
($\alpha=e,\mu,\tau$; $\beta=e,\mu$)
\begin{align}
&\sum_{A,B}({\cal M}_{H_C}^{-1})_{AB}\left\{(Y_R^A)_{\tau t}(\ov{Y}^B_R)_{ud}-(Y_R^A)_{\tau u}(\ov{Y}^B_R)_{td}\right\}
\nn\\
&=\frac{1}{M_{H_C}}
\left[
a\left\{(Y_{10})_{\tau_R t_R}(Y_{10})_{u_Rd_R}-(Y_{10})_{\tau_R u_R}(Y_{10})_{t_Rd_R}\right\}
+b\left\{(Y_{10})_{\tau_R t_R}(Y_{126})_{u_Rd_R}-(Y_{10})_{\tau_R u_R}(Y_{126})_{t_Rd_R}\right\}\right.
\nn\\
&+c\left\{(Y_{10})_{\tau_R t_R}(Y_{120})_{u_Rd_R}-(Y_{10})_{\tau_R u_R}(Y_{120})_{t_Rd_R}\right\}
\nn\\
&+d\left\{(Y_{126})_{\tau_R t_R}(Y_{10})_{u_Rd_R}-(Y_{126})_{\tau_R u_R}(Y_{10})_{t_Rd_R}\right\}
+e\left\{(Y_{126})_{\tau_R t_R}(Y_{126})_{u_Rd_R}-(Y_{126})_{\tau_R u_R}(Y_{126})_{t_Rd_R}\right\}
\nn\\
&+f\left\{(Y_{126})_{\tau_R t_R}(Y_{120})_{u_Rd_R}-(Y_{126})_{\tau_R u_R}(Y_{120})_{t_Rd_R}\right\}
\nn\\
&+g\left\{(Y_{120})_{\tau_R t_R}(Y_{10})_{u_Rd_R}-(Y_{120})_{\tau_R u_R}(Y_{10})_{t_Rd_R}\right\}
+h\left\{(Y_{120})_{\tau_R t_R}(Y_{126})_{u_Rd_R}-(Y_{120})_{\tau_R u_R}(Y_{126})_{t_Rd_R}\right\}
\nn\\
&\left.+j\left\{(Y_{120})_{\tau_R t_R}(Y_{120})_{u_Rd_R}-(Y_{120})_{\tau_R u_R}(Y_{120})_{t_Rd_R}\right\}\right],
\nn\\
\label{yyr1}
\\
&\sum_{A,B}({\cal M}_{H_C}^{-1})_{AB}\left\{(Y_R^A)_{\tau t}(\ov{Y}^B_R)_{us}-(Y_R^A)_{\tau u}(\ov{Y}^B_R)_{ts}\right\}
={\rm (Above \ expression \ with \ exchange} \ d_R\leftrightarrow s_R),
\label{yyr2}
\\
&\sum_{A,B}({\cal M}_{H_C}^{-1})_{AB}\left\{(Y_L^A)_{ud}(\ov{Y}^B_L)_{s\alpha}-(Y_L^A)_{ds}(\ov{Y}^B_L)_{u\alpha}\right\}
\nn\\
&=\frac{1}{M_{H_C}}
\left[
a\left\{(Y_{10})_{u_L d_L}(Y_{10})_{s_L\alpha_L}-(Y_{10})_{d_L s_L}(Y_{10})_{u_L\alpha_L}\right\}
+b\left\{(Y_{10})_{u_L d_L}(Y_{126})_{s_L\alpha_L}-(Y_{10})_{d_L s_L}(Y_{126})_{u_L\alpha_L}\right\}\right.
\nn\\
&+c\left\{(Y_{10})_{u_L d_L}(Y_{120})_{s_L\alpha_L}-(Y_{10})_{d_L s_L}(Y_{120})_{u_L\alpha_L}\right\}
\nn\\
&+d\left\{(Y_{126})_{u_L d_L}(Y_{10})_{s_L\alpha_L}-(Y_{126})_{d_L s_L}(Y_{10})_{u_L\alpha_L}\right\}
+e\left\{(Y_{126})_{u_L d_L}(Y_{126})_{s_L\alpha_L}-(Y_{126})_{d_L s_L}(Y_{126})_{u_L\alpha_L}\right\}
\nn\\
&\left.+f\left\{(Y_{126})_{u_L d_L}(Y_{120})_{s_L\alpha_L}-(Y_{126})_{d_L s_L}(Y_{120})_{u_L\alpha_L}\right\}\right],
\nn\\
\label{yyl1}
\\
&\sum_{A,B}({\cal M}_{H_C}^{-1})_{AB}\left\{(Y_L^A)_{us}(\ov{Y}^B_L)_{d\alpha}-(Y_L^A)_{ds}(\ov{Y}^B_L)_{u\alpha}\right\}
={\rm (Above \ expression \ with \ exchange} \ d_L\leftrightarrow s_L),
\label{yyl2}
\\
&\sum_{A,B}({\cal M}_{H_C}^{-1})_{AB}\left\{(Y_L^A)_{us}(\ov{Y}^B_L)_{u\beta}-(Y_L^A)_{uu}(\ov{Y}^B_L)_{s\beta}\right\}
\nn\\
&=\frac{1}{M_{H_C}}
\left[
a\left\{(Y_{10})_{u_L s_L}(Y_{10})_{u_L\beta_L}-(Y_{10})_{u_L u_L}(Y_{10})_{s_L\beta_L}\right\}
+b\left\{(Y_{10})_{u_L s_L}(Y_{126})_{u_L\beta_L}-(Y_{10})_{u_L u_L}(Y_{126})_{s_L\beta_L}\right\}\right.
\nn\\
&+c\left\{(Y_{10})_{u_L s_L}(Y_{120})_{u_L\beta_L}-(Y_{10})_{u_L u_L}(Y_{120})_{s_L\beta_L}\right\}
\nn\\
&+d\left\{(Y_{126})_{u_L s_L}(Y_{10})_{u_L\beta_L}-(Y_{126})_{u_L u_L}(Y_{10})_{s_L\beta_L}\right\}
+e\left\{(Y_{126})_{u_L s_L}(Y_{126})_{u_L\beta_L}-(Y_{126})_{u_L u_L}(Y_{126})_{s_L\beta_L}\right\}
\nn\\
&\left.+f\left\{(Y_{126})_{u_L s_L}(Y_{120})_{u_L\beta_L}-(Y_{126})_{u_L u_L}(Y_{120})_{s_L\beta_L}\right\}\right],
\nn\\
\label{yyl3}
\end{align}
 where $M_{H_C}$ is the scale of the singular values of ${\cal M}_{H_C}$, and $a,b,c,d,e,f,g,h,j$ are $O(1)$ numbers
 determined from ${\cal M}_{H_C}$ as Ref.~\cite{Fukuyama:2004xs}-\cite{Bajc:2005qe}.
Here $(Y_{10})_{\tau_Rt_R}$ denotes the component of $Y_{10}$ in the term $(Y_{10})_{ij}\,\Psi_i H \Psi_j$
 that involves the right-handed tau lepton component of $\Psi_i$ and the right-handed top quark component of $\Psi_j$.
Other symbols are defined analogously.
\\

\section{Conditions for a Texture of the Yukawa Coupling Matrices suppressing Dimension-five Proton Decay}
\label{section-identification}

We identify those components of the Yukawa coupling matrices $Y_{10},Y_{126},Y_{120}$
 that are involved in dimension-five proton decay, namely, appear in Eqs.~(\ref{yyr1})-(\ref{yyl3}),
 and that can be on the order of the up quark Yukawa coupling $y_u$.
Here the up quark Yukawa coupling is considered as the smallest scale of the components of the Yukawa coupling matrices
 because it is a specially small Yukawa coupling in the SUSY $SO(10)$ GUT where $\tan\beta\sim50$.
That the components identified above be on the order of the up quark Yukawa coupling, is the desired conditions for a texture suppressing dimension-five proton decay.
\\

\begin{itemize}

\item
We focus on the first term in each \{...\} of Eqs.~(\ref{yyr1}),(\ref{yyr2}).
Components $(Y_{10})_{\tau_R t_R}$ and $(Y_{126})_{\tau_R t_R}$ are almost $(3,3)$-components of the symmetric matrices $Y_{10},Y_{126}$
 and hence {\bf always} on the order of the top quark Yukawa coupling $y_t$.
For other components, at most two of $(Y_{10})_{u_R d_R}$, $(Y_{126})_{u_R d_R}$, $(Y_{120})_{u_R d_R}$
 and at most two of $(Y_{10})_{u_R s_R}$, $(Y_{126})_{u_R s_R}$, $(Y_{120})_{u_R s_R}$
 can be on the order of the up quark Yukawa coupling $y_u$.
However, all of them cannot be so because of the following equalities that result from Eq.~(\ref{yd}):
\begin{align}
(Y_{10})_{u_R d_R}+(Y_{126})_{u_R d_R}+(Y_{120})_{u_R d_R} \ &= \ \frac{1}{r_1}(Y_d)_{u_R d_R}
\nn\\
&\simeq \ \frac{y_t}{y_b}y_d\times (d_L\mathchar`-u_R \ {\rm part \ of \ the \ mixing \ matrix}),
 \label{urdr}
\\
(Y_{10})_{u_R s_R}+(Y_{126})_{u_R s_R}+(Y_{120})_{u_R s_R} \ &= \ \frac{1}{r_1}(Y_d)_{u_R s_R}
\nn\\
&\simeq \ \frac{y_t}{y_b}y_s\times (s_L\mathchar`-u_R \ {\rm part \ of \ the \ mixing \ matrix}),
 \label{ursr}
\end{align}
 where $r_1$ is estimated to be $y_b/y_t$ so that the ratio of the top and bottom quark Yukawa couplings is reproduced.
$d_L\mathchar`-u_R \ {\rm part \ of \ the \ mixing \ matrix}$ is estimated to be about 1 and we get
 $(Y_{10})_{u_R d_R}+(Y_{126})_{u_R d_R}+(Y_{120})_{u_R d_R}\simeq\frac{y_t}{y_b}y_d$, which is much greater than $y_u$.
Also, $s_L\mathchar`-u_R \ {\rm part \ of \ the \ mixing \ matrix}$
 is estimated to be the Cabibbo angle $\lambda=0.22$ and hence
 we get $(Y_{10})_{u_R s_R}+(Y_{126})_{u_R s_R}+(Y_{120})_{u_R s_R}\simeq0.22\times\frac{y_t}{y_b}y_s$,
 which is much greater than $y_u$.
Therefore, to make the entire Eqs.~(\ref{yyr1}),(\ref{yyr2}) proportional to $y_u$,
 we have to tune the colored Higgs mass matrix such that some of $a,b,c,d,e,f,g,h,j$ are much smaller than 1.
The most economical choice is to tune the colored Higgs mass matrix to make
 \bea
c=f=0
 \label{coloredHiggsTexture}
 \eea
  and at the same time consider the following texture:
\bea
&&(Y_{10})_{u_R d_R}=O(y_u), \ \ \ \ \ (Y_{126})_{u_R d_R}=O(y_u), \ \ \ \ \ (Y_{120})_{u_R d_R}\simeq\frac{y_t}{y_b}y_d,
\label{texture1}
\\
&&(Y_{10})_{u_R s_R}=O(y_u), \ \ \ \ \ (Y_{126})_{u_R s_R}=O(y_u), \ \ \ \ \ (Y_{120})_{u_R s_R}\simeq0.22\times\frac{y_t}{y_b}y_s.
\label{texture2}
\eea

The terms $j(Y_{120})_{\tau_R t_R}(Y_{120})_{u_R d_R}$ and $j(Y_{120})_{\tau_R t_R}(Y_{120})_{u_R s_R}$ appear to be not proportional to $y_u$.
However, since $Y_{120}$ is antisymmetric in the flavor space, $(Y_{120})_{\tau_R t_R}$, being a nearly diagonal component, is so small that
 the above terms are suppressed compared to the other remaining terms.
Note that $(Y_{120})_{\tau_R t_R}$ is not exactly 0 because the flavor basis of the isospin-singlet charged leptons $(e_R,\mu_R,\tau_R)$
 is not identical with that of the isospin-singlet up-type quarks $(u_R,c_R,t_R)$. However, these flavor bases are close and thus $(Y_{120})_{\tau_R t_R}$ is suppressed.

\item
We focus on the second term in each \{...\} of Eq.~(\ref{yyr1}).
Each term is estimated to be $\sin^2\theta^{\rm ckm}_{13}\, y_t^2$ ($\theta^{\rm ckm}_{13}$ is the $(1,3)$-mixing angle of CKM matrix),
 since each term contains two 1st-3rd generation flavor mixings.
The value of $\sin^2\theta^{\rm ckm}_{13}\, y_t^2$ is numerically close to $y_u\,y_t$.
Thus, these terms are {\bf always} on the same order as first terms
 $(Y_{10})_{\tau_R t_R}(Y_{10})_{u_R d_R}$, $(Y_{10})_{\tau_R t_R}(Y_{126})_{u_R s_R}$, $(Y_{126})_{\tau_R t_R}(Y_{10})_{u_R d_R}$, $(Y_{126})_{\tau_R t_R}(Y_{126})_{u_R s_R}$
 with the texture of Eq.~(\ref{texture1}).

Likewise the second term in each \{...\} of Eq.~(\ref{yyr2}) is estimated to be $\sin\theta^{\rm ckm}_{13}\sin\theta^{\rm ckm}_{23}\,y_t^2$.
These terms contribute to the proton decay amplitude
 {\bf always} by a similar amount to the second terms in \{...\}'s of Eq.~(\ref{yyr1})
 because they enter the proton decay amplitude in the form
 $V_{ts}^{\rm ckm}(Y_A)_{\tau_R u_R}(Y_B)_{t_R d_R}$ and $V_{td}^{\rm ckm}(Y_A)_{\tau_R u_R}(Y_B)_{t_R s_R}$
 and the CKM matrix components satisfy $|V_{ts}^{\rm ckm}|\sim\sin\theta^{\rm ckm}_{23}$ and 
 $|V_{td}^{\rm ckm}|\sim\sin\theta^{\rm ckm}_{13}$.

\item
We proceed to Eqs.~(\ref{yyl1}),(\ref{yyl2}).
It is impossible to suppress $(Y_{A})_{s_L \alpha_L}$, $(Y_{A})_{d_L \alpha_L}$, $(Y_{A})_{u_L \alpha_L}$ for {\bf all flavors} $\alpha=e,\mu,\tau$ to the order of $y_u$.
Therefore, we do not consider a texture where some of them are $O(y_u)$.
For other components, at least one of $(Y_{10})_{u_L s_L}$, $(Y_{10})_{d_L s_L}$, $(Y_{126})_{u_L s_L}$, $(Y_{126})_{d_L s_L}$ is on the order of
 $|V^{\rm ckm}_{cd}|\frac{y_t}{y_b}y_s$ because of two equalities below,
\begin{align}
&(Y_{10})_{s_L c_L}+(Y_{126})_{s_L c_L}+(Y_{120})_{s_L c_L} \ \simeq \ \frac{y_t}{y_b}y_s
\times (c_L\mathchar`-s_R \ {\rm part \ of \ the \ mixing \ matrix)},
\label{clsl}
\\
&(Y_{10})_{d_L s_L}+(Y_{126})_{d_L s_L}-V^{\rm ckm}_{ud} \left\{ (Y_{10})_{u_L s_L}+(Y_{126})_{u_L s_L} \right\}
\nn\\
&\ \ \ \ \ \ \ \ =\ V^{\rm ckm}_{cd} \left\{ (Y_{10})_{c_L s_L}+(Y_{126})_{c_L s_L} \right\}
+V^{\rm ckm}_{td} \left\{ (Y_{10})_{t_L s_L}+(Y_{126})_{t_L s_L} \right\},
\label{dlsl}
\end{align}
 and by the facts that $c_L\mathchar`-s_R \ {\rm part \ of \ the \ mixing \ matrix}$ is about 1 because $c_L$ and $s_R$ are 2nd generation flavors,
 and that $(Y_{120})_{s_L c_L}$ is suppressed compared to $(Y_{10})_{c_L s_L},(Y_{126})_{c_L s_L}$ because it is nearly $(2,2)$-component of the antisymmetric matrix $Y_{120}$.\footnote{
The situation that the term $V^{\rm ckm}_{td} \left\{ (Y_{10})_{t_L s_L}+(Y_{126})_{t_L s_L} \right\}$
 cancels the term $V^{\rm ckm}_{cd} \left\{ (Y_{10})_{c_L s_L}+(Y_{126})_{c_L s_L} \right\}$ is incompatible with the correct quark Yukawa couplings.}
As a result, at least one of $(Y_{10})_{u_L s_L}$, $(Y_{10})_{d_L s_L}$, $(Y_{126})_{u_L s_L}$, $(Y_{126})_{d_L s_L}$ cannot be on the order of $y_u$.
Still, it is possible to make the entire Eqs.~(\ref{yyl1}),(\ref{yyl2}) proportional to $y_u$ 
 by tuning the colored Higgs mass matrix such that $a,b,d,e$ satisfy
\bea
a\, (Y_{10})_{d_L s_L}+d\, (Y_{126})_{d_L s_L}=0, \ \ \ \ \ \ \ b\, (Y_{10})_{d_L s_L}+e\, (Y_{126})_{d_L s_L}=0
\eea
 and at the same time considering the following texture:
\bea
(Y_{10})_{u_L s_L}=O(y_u), \ \ \ \ \ (Y_{126})_{u_L s_L}=O(y_u).
\eea

\item
Finally, we focus on Eq.~(\ref{yyl3}).
The only components that do not appear in Eqs.~(\ref{yyl1}),(\ref{yyl2}) are $(Y_{10})_{u_L u_L}$, $(Y_{126})_{u_L u_L}$,
 and there is no obstacle in considering the following texture:
\bea
(Y_{10})_{u_L u_L}=O(y_u), \ \ \ \ \ (Y_{126})_{u_L u_L}=O(y_u).
\eea

\end{itemize}

To summarize, dimension-five proton decay is suppressed if the components below are all on the order of the up quark Yukawa coupling $y_u$,
\bea
&&(Y_{10})_{u_R d_R}, \ (Y_{126})_{u_R d_R}, \ (Y_{10})_{u_R s_R}, \ (Y_{126})_{u_R s_R},
\nn\\
&&(Y_{10})_{u_L d_L}, \ (Y_{126})_{u_L d_L}, \ (Y_{10})_{u_L s_L}, \ (Y_{126})_{u_L s_L}, \ (Y_{10})_{u_L u_L}, \ (Y_{126})_{u_L u_L},
\label{tobeyu}
\eea
 and at the same time the colored Higgs mass matrix is tuned such that $a,b,c,d,e,f,g,h,j$ in Eqs.~(\ref{yyr1})-(\ref{yyl3}) satisfy
\bea
&&c=f=0,
\label{coloredhiggstex1}\\
&&a\, (Y_{10})_{d_L s_L}+d\, (Y_{126})_{d_L s_L}=0, \ \ \ \ \ \ b\, (Y_{10})_{d_L s_L}+e\, (Y_{126})_{d_L s_L}=0.
\label{coloredhiggstex2}
\eea
That the components of Eq.~(\ref{tobeyu}) be on the order of the up quark Yukawa coupling, is the desired conditions for a texture of the Yukawa coupling matrices suppressing dimension-five proton decay.

For reference, below we summarize the estimates on crucial Yukawa coupling components involved in dimension-five proton decay other than Eq.~(\ref{tobeyu}),
\begin{align}
&(Y_{10})_{\tau_R t_R} \sim (Y_{126})_{\tau_R t_R} \sim y_t, \ \ \ \ \ \ \ \ \ \ \ (Y_{120})_{\tau_R t_R} \ll y_t,
\nn\\
&(Y_{120})_{u_R d_R} \sim \frac{y_t}{y_b}y_d, \ \ \ \ \ \ \ \ \ \ \ \ \ \ \ \ \ \ \ \ \ \ (Y_{120})_{u_R s_R} \sim \lambda\frac{y_t}{y_b}y_s,
\nn\\
&(Y_A)_{\tau_R u_R} \sim \ (Y_A)_{t_R d_R} \sim y_t \, \sin\theta_{13}^{\rm ckm} \ \ \ {\rm for} \ \ \ A=10,126,120,
\nn\\
&(Y_{10})_{d_L s_L} \sim (Y_{126})_{d_L s_L} \sim \lambda\frac{y_t}{y_b}y_s,
\end{align}
 where $\lambda=0.22$ and $\sin\theta_{13}^{\rm ckm}=0.004$.

We comment on nucleon decay modes other than Eqs.~(\ref{ptoknutau})-(\ref{ptok0mu}).
The partial widths of the $N\to \pi \beta^+$ and $p\to \eta \beta^+$ decays ($\beta=e,\mu$)
 involve the same Yukawa coupling components as the $p\to K^0\beta^+$ except that $s_L$ is replaced by $d_L$.
Hence, once we consider the texture where $(Y_{10})_{u_L u_L}$, $(Y_{126})_{u_L u_L}$ are on the order of the up quark Yukawa coupling,
 the $N\to \pi \beta^+$ and $p\to \eta \beta^+$ decays are also suppressed.
Constraints on the rest of the decay modes are relatively weak~\cite{ParticleDataGroup:2022pth} and are not in tension with the SUSY $SO(10)$ GUT.
\\

\section{Texture of the Colored Higgs Mass Matrix}
\label{section-coloredhiggs}

We present a texture of the colored Higgs mass matrix that gives $c=f=0$ and $a/d=b/e$.
Here $a/d=b/e$ is a necessary condition for Eq.~(\ref{coloredhiggstex2}).
We utilize the result of Ref.~\cite{Fukuyama:2004ps}, and use the same notation of fields, coupling constants and VEVs except that {\bf 120} field is written as $\Sigma$ in our paper.
The definitions of the couplings, coupling constants and masses are summarized in Appendix~A.

The desired texture of the colored Higgs mass matrix is obtained from the following relations of the coupling constants and VEVs:
\begin{align}
\lambda_{18} \ &= \ 0, \ \ \ \ \ 
\lambda_{20} \ = \ 0, \ \ \ \ \
\frac{\lambda_{21}}{\lambda_{19}}\ = \ 3\frac{\lambda_{17}}{\lambda_{16}},
\nn\\
i\,A_1 \ &= \ -\frac{1}{6}\frac{\lambda_{21}}{\lambda_{19}}\Phi_3, \  \ \ \ \ 
i\,A_2 \ = \ -\frac{\sqrt{3}}{6}\frac{\lambda_{21}}{\lambda_{19}}\Phi_2.
\label{texture}
\end{align}
The above relations are obtained by fine-tuning, which is natural at the quantum level thanks to the non-renormalization theorem.
The VEV configuration of Eq.~(\ref{texture}) can be consistent with the $F$-flatness conditions of the six SM-gauge-singlet components $\Phi_1,\Phi_2,\Phi_3,A_1,A_2,v_R$
 when six model parameters, for example $\lambda_5/\lambda_2,\,\lambda_6/\lambda_2,\,\lambda_7/\lambda_2,\,m_1,m_2,m_4$, are tuned appropriately.

When Eq.~(\ref{texture}) holds, the colored Higgs mass matrix has the following texture:
\begin{align}
   &W_{\rm colored\,Higgs} \supset \begin{pmatrix} 
      H^{(\ov{3},1,\frac{1}{3})}  &  \ov{\Delta}^{(\ov{3},1,\frac{1}{3})}_{(6,1,1)}  & \Delta^{(\ov{3},1,\frac{1}{3})}_{(6,1,1)}  & \Delta^{(\ov{3},1,\frac{1}{3})}_{(\ov{10},1,3)}
      & \Phi^{(\ov{3},1,\frac{1}{3})} & \Sigma^{(\ov{3},1,\frac{1}{3})}_{(6,1,3)} & \Sigma^{(\ov{3},1,\frac{1}{3})}_{(\ov{10},1,1)}
   \end{pmatrix}{\cal M}_{H_C}
   \begin{pmatrix} 
H^{(3,1,-\frac{1}{3})}  \\  \Delta^{(3,1,-\frac{1}{3})}_{(6,1,1)}  \\ \ov{\Delta}^{(3,1,-\frac{1}{3})}_{(6,1,1)} \\ \ov{\Delta}^{(3,1,-\frac{1}{3})}_{(10,1,3)}
      \\ \Phi^{(3,1,-\frac{1}{3})} \\ \Sigma^{(3,1,-\frac{1}{3})}_{(6,1,3)} \\ \Sigma^{(3,1,-\frac{1}{3})}_{(10,1,1)}
   \end{pmatrix},
\end{align}
where
\begin{align}
&{\cal M}_{H_C}
=\begin{pmatrix} 
      m_3 &\frac{ \lambda_3\Phi_2}{\sqrt{30}}-\frac{ \lambda_3\Phi_1}{\sqrt{10}} & -\frac{\lambda_4\Phi_1}{\sqrt{10}}-\frac{\lambda_4\Phi_2}{\sqrt{30}}
      & -\sqrt{\frac{2}{15}}\lambda_4\Phi_3 & \frac{\lambda_4\ov{v}_R}{\sqrt{5}} & 0 & 0 \\
      \frac{\lambda_4\Phi_2}{\sqrt{30}}-\frac{\lambda_4\Phi_1}{\sqrt{10}} & m_2+i\frac{\lambda_{21}}{\lambda_{19}}\frac{\lambda_6\Phi_2}{30\sqrt{2}} & 0 & 0 &0 &0 & 0 \\
      -\frac{\lambda_3\Phi_1}{\sqrt{10}}-\frac{\lambda_3\Phi_2}{\sqrt{30}} & 0 & m_2-i\frac{\lambda_{21}}{\lambda_{19}}\frac{\lambda_6\Phi_2}{30\sqrt{2}} &\frac{\lambda_2\Phi_3}{15\sqrt{2}} & -\frac{\lambda_2\ov{v}_R}{10\sqrt{3}} & 0 & 0 \\
      -\sqrt{\frac{2}{15}}\lambda_3\Phi_3 & 0 & \frac{\lambda_2\Phi_3}{15\sqrt{2}} & m_{66}  & -\frac{\lambda_2\ov{v}_R}{5\sqrt{6}} & 0 & 0 \\
      \frac{\lambda_3v_R}{\sqrt{5}} & 0 & -\frac{\lambda_2v_R}{10\sqrt{3}} & -\frac{\lambda_2v_R}{5\sqrt{6}} & m_{77} & 0 & 0 \\
      -\lambda_{17}\frac{\Phi_3}{\sqrt{3}} & 0 & \lambda_{21}\frac{\Phi_3}{6\sqrt{5}} & \lambda_{21}\frac{\Phi_2}{3\sqrt{5}} & \frac{\lambda_{21}\ov{v}_R}{2\sqrt{15}} & m_{22} & \frac{2\lambda_{15}\Phi_3}{9} \\
      -\sqrt{\frac{2}{3}}\lambda_{17}\Phi_2 & 0 & \lambda_{21}\frac{\Phi_2}{3\sqrt{10}} & \lambda_{21}\frac{\Phi_3}{3\sqrt{10}} &
 \frac{\lambda_{21}\ov{v}_R}{2\sqrt{15}} & \frac{2\lambda_{15}\Phi_3}{9} & m_{33}
   \end{pmatrix}
 \label{texture-coloredhiggs}
 \\
&m_{66}= m_2+\lambda_2(\frac{\Phi_1}{10\sqrt{6}}+\frac{\Phi_2}{30\sqrt{2}})-i\frac{\lambda_{21}}{\lambda_{19}}\frac{\lambda_6\Phi_2}{30\sqrt{2}}
\\
&m_{77} = m_1+\lambda_1(\frac{\Phi_1}{\sqrt{6}}+\frac{\Phi_2}{3\sqrt{2}}+\frac{2\Phi_3}{3})-i\frac{\lambda_{21}}{\lambda_{19}}\frac{\sqrt{2}\lambda_7\Phi_2}{15}
\\
&m_{22} = m_6+\frac{1}{3}\sqrt{\frac{2}{3}}\lambda_{15}\Phi_1
\\
&m_{33} = m_6+\frac{\sqrt{2}}{9}\lambda_{15}\Phi_2.
\end{align}
The dimension-five operators responsible for proton decay Eq.~(\ref{w5}) satisfy at the scale of the colored Higgs mass,
\begin{align}
W_5=&-\begin{pmatrix} 
      (\tilde{Y}_{10})_{ij} &  0  &  (\tilde{Y}_{126})_{ij} & (\tilde{Y}_{126})_{ij}
      & 0 & (\tilde{Y}_{120})_{ij} & (\tilde{Y}_{120})_{ij}
   \end{pmatrix}{\cal M}_{H_C}^{-1}
   \begin{pmatrix} 
(\tilde{Y}_{10})_{kl}  \\  (\tilde{Y}_{126})_{kl}  \\ 0 \\ 0 \\ 0 \\ (\tilde{Y}_{120})_{kl} \\ (\tilde{Y}_{120})_{kl}
   \end{pmatrix}E_i^cU_j^cU_k^cD_l^c
\nonumber\\
&-\begin{pmatrix} 
      (\tilde{Y}_{10})_{ij} &  0  &  (\tilde{Y}_{126})_{ij} & (\tilde{Y}_{126})_{ij}
      & 0 & 0 & 0
   \end{pmatrix}{\cal M}_{H_C}^{-1}
   \begin{pmatrix} 
(\tilde{Y}_{10})_{kl}  \\  (\tilde{Y}_{126})_{kl}  \\ 0 \\ 0 \\ 0 \\ (\tilde{Y}_{120})_{kl} \\ (\tilde{Y}_{120})_{kl}
   \end{pmatrix}(Q_iQ_j)(Q_kL_l),
\end{align}
 where the inverse of the colored Higgs mass ${\cal M}_{H_C}^{-1}$ has the following properties resulting from Eq.~(\ref{texture-coloredhiggs}):
\begin{itemize}

\item
The upper-right $5\times 2$ part of ${\cal M}_{H_C}^{-1}$ is zero because the upper-right $5\times 2$ part of ${\cal M}_{H_C}$ is zero.
Thus, the coefficients proportional to $(\tilde{Y}_{10})_{ij}(\tilde{Y}_{120})_{kl}$ or $(\tilde{Y}_{126})_{ij}(\tilde{Y}_{120})_{kl}$ are zero, namely,
 we get $c=f=0$ in Eqs.~(\ref{yyr1})-(\ref{yyl3}).

\item
The upper-left $5\times5$ part of ${\cal M}_{H_C}^{-1}$ is given by the inverse of the upper-left $5\times5$ part of ${\cal M}_{H_C}$, since the upper-right $5\times 2$ part of ${\cal M}_{H_C}$ is zero.
Then the equalities $({\cal M}_{H_C})_{32}=({\cal M}_{H_C})_{42}=({\cal M}_{H_C})_{52}=0$
 lead to the relation $({\cal M}_{H_C}^{-1})_{11}:({\cal M}_{H_C}^{-1})_{31}:({\cal M}_{H_C}^{-1})_{41}=
({\cal M}_{H_C}^{-1})_{12}:({\cal M}_{H_C}^{-1})_{32}:({\cal M}_{H_C}^{-1})_{42}$.
Since the coefficients proportional to $(\tilde{Y}_{10})_{ij}(\tilde{Y}_{10})_{kl}$, $(\tilde{Y}_{10})_{ij}(\tilde{Y}_{126})_{kl}$, $(\tilde{Y}_{126})_{ij}(\tilde{Y}_{10})_{kl}$ or $(\tilde{Y}_{126})_{ij}(\tilde{Y}_{126})_{kl}$ are given by
\begin{align}
&({\cal M}_{H_C}^{-1})_{11}\ (\tilde{Y}_{10})_{ij}(\tilde{Y}_{10})_{kl}+\left\{({\cal M}_{H_C}^{-1})_{31}+({\cal M}_{H_C}^{-1})_{41}\right\}(\tilde{Y}_{126})_{ij}(\tilde{Y}_{10})_{kl}
\nonumber\\
&+({\cal M}_{H_C}^{-1})_{12}\ (\tilde{Y}_{10})_{ij}(\tilde{Y}_{126})_{kl}+\left\{({\cal M}_{H_C}^{-1})_{32}+({\cal M}_{H_C}^{-1})_{42}\right\}(\tilde{Y}_{126})_{ij}(\tilde{Y}_{126})_{kl},
\end{align}
the equality $({\cal M}_{H_C}^{-1})_{11}/\left\{({\cal M}_{H_C}^{-1})_{31}+({\cal M}_{H_C}^{-1})_{41}\right\}
=({\cal M}_{H_C}^{-1})_{12}/\left\{({\cal M}_{H_C}^{-1})_{32}+({\cal M}_{H_C}^{-1})_{42}\right\}$
 leads to the desired relation $a/d=b/e$ in Eqs.~(\ref{yyr1})-(\ref{yyl3}).

\end{itemize}

We still have to check that Eq.~(\ref{texture}) is consistent with the situation that $a/d$ satisfies $a\, (Y_{10})_{d_L s_L}+d\, (Y_{126})_{d_L s_L}=0$,
 that only one pair of (${\bf 1}$, ${\bf2}$, $\pm\frac{1}{2}$) fields (corresponding to the MSSM Higgs fields) have almost zero mass compared to the GUT scale,
 and that $r_1,r_2,r_3,r_e,r_D$ take values that reproduce the realistic quark and lepton masses and flavor mixings.
With Eq.~(\ref{texture}), and with the tuning of $\lambda_5/\lambda_2,\,\lambda_6/\lambda_2,\,\lambda_7/\lambda_2,\,m_1,m_2,m_4$ to satisfy the $F$-flatness conditions,
 the colored Higgs mass matrix and the mass matrix of the $({\bf 1},{\bf 2},\pm\frac{1}{2})$ components
 still have free parameters $\Phi_1,\Phi_2,\Phi_3,v_R$, $m_3$, $\lambda_{21}/\lambda_{19},\,\lambda_{21}/\lambda_2$, $\lambda_1/\lambda_2,\,\lambda_3/\lambda_2,\,\lambda_4/\lambda_2,\,\lambda_{15}/\lambda_2,\,\lambda_{17}/\lambda_2$.
These free parameters are sufficient to make one pair of (${\bf 1}$, ${\bf2}$, $\pm\frac{1}{2}$) fields nearly massless and realize
 any values of $a/d$, $r_1,r_2,r_3,r_e,r_D$.
\\

\section{Yukawa Coupling Matrices satisfying the Conditions}
\label{section-fitting}

\subsection{Procedures of the Analysis}
\label{section-fitting1}

We will obtain specific Yukawa coupling matrices $Y_{10}$, $Y_{126}$, $Y_{120}$ which give the realistic quark and lepton masses and flavor mixings and which satisfy the conditions found in Section~\ref{section-identification} that
 the components of Eq.~(\ref{tobeyu}) be on the order of the up quark Yukawa coupling $y_u$.
To this end, we fit the experimental values of the quark and lepton masses and flavor mixings
 with $Y_{10}$, $Y_{126}$, $Y_{120}$ and numbers $r_1,r_2,r_3,r_e,r_D$ based on Eqs.~(\ref{yu})-(\ref{cnu})
 and at the same time minimize the following quantity:
\begin{align}
&\frac{1}{y_u^2}\sum_{A=10,126} \ 
\left|(Y_A)_{u_R d_R}\right|^2+\left|(Y_A)_{u_R s_R}\right|^2+\left|(Y_A)_{u_L d_L}\right|^2+\left|(Y_A)_{u_L s_L}\right|^2+\left|(Y_A)_{u_L u_L}\right|^2.
\label{tobeminimized}
\end{align}
The procedures are as follows:
\\

We adopt the following experimental values of the quark and charged lepton masses, quark flavor mixings and gauge coupling constants:
We use the results of lattice calculations of the individual up and down quark masses, the strange quark mass, the charm quark mass and the bottom quark mass in $\ov{{\rm MS}}$ scheme reviewed in Ref.~\cite{FlavourLatticeAveragingGroupFLAG:2021npn}, 
 which read $m_u(2~{\rm GeV})=2.14(8)~{\rm MeV}$, $m_d(2~{\rm GeV})=4.70(5)~{\rm MeV}$~\cite{FermilabLattice:2018est,Giusti:2017dmp}, 
 $m_s(2~{\rm GeV})=93.40(57)~{\rm MeV}$~\cite{FermilabLattice:2018est,EuropeanTwistedMass:2014osg,Lytle:2018evc,Chakraborty:2014aca}, 
 $m_c(3~{\rm GeV})=0.988(11)~{\rm GeV}$~\cite{FermilabLattice:2018est,EuropeanTwistedMass:2014osg,Chakraborty:2014aca,Alexandrou:2014sha,Hatton:2020qhk}, 
 $m_b(m_b)=4.203(11)~{\rm GeV}$~\cite{FermilabLattice:2018est,Chakraborty:2014aca,Hatton:2021syc,Colquhoun:2014ica,ETM:2016nbo,Gambino:2017vkx}.
We use the top quark pole mass measured by CMS in Ref.~\cite{Sirunyan:2019zvx}, which reads $M_t=170.5(8)$~GeV.
We calculate the CKM mixing angles and CP phase from the Wolfenstein parameters in Ref.~\cite{ckmfitter}.
The lepton pole masses and $W$, $Z$, Higgs boson pole masses are taken from Particle Data Group~\cite{ParticleDataGroup:2022pth},
 and the QCD and QED gauge coupling constants in 5-quark-flavor QCD$\times$QED theory are fixed as $\alpha_s^{(5)}(M_Z)=0.1181$ and $\alpha^{(5)}(M_Z)=1/127.95$.
The above data are translated into the values of the quark and lepton Yukawa coupling matrices and gauge coupling constants
 at scale $\mu=M_Z$ in $\ov{{\rm MS}}$ scheme
 with the help of the code~\cite{code} based on Refs.~\cite{Jegerlehner:2001fb}-\cite{threshold}.

We calculate the two-loop RG equations~\cite{Machacek:1983tz}-\cite{Machacek:1984zw} of SM from scale $\mu=M_Z$ to the soft SUSY breaking scale $\mu_{\rm SUSY}$.
The results are matched to the Yukawa coupling matrices and gauge couplings of MSSM in $\ov{{\rm DR}}$ scheme.
Here the one-loop threshold corrections of SUSY particles, which are important for the down-type quark and charged lepton Yukawa couplings as $\tan\beta$ is large,
 are included as Ref.~\cite{Blazek:1995nv}.
Then we calculate the two-loop RG equations of MSSM from scale $\mu_{\rm SUSY}$ to the GUT scale $\mu_{\rm GUT}$.
We assume a degenerate SUSY particle mass spectrum where the pole masses of SUSY particles and $\tan\beta$ are given by
\bea
&&m_{\rm sfermion}=m_{H^0}=m_{H^\pm}=m_A=1500~{\rm TeV},
\nn\\
&&|M_{\widetilde{g}}|=|M_{\widetilde{W}}|=|\mu_H|=1500~{\rm TeV},
\ \ \ \ \ \ \tan\beta = 50
\label{massspectrum}
\eea
 and all the $A$-terms are 0.
We set $\mu_{\rm SUSY}=1500$~TeV and $\mu_{\rm GUT}=2\cdot10^{16}$~GeV.
The values of the Yukawa coupling matrices at scale $\mu=\mu_{\rm GUT}$ are shown in Table~\ref{values},
 in the form of the singular values of the matrices and the parameters of the CKM matrix at this scale.
The errors of the quark Yukawa couplings, propagated from the experimental errors of the corresponding masses,
 and the maximal errors of the CKM parameters, obtained by assuming maximal correlation of the experimental errors of the Wolfenstein parameters, are also displayed.
\begin{table}[H]
\begin{center}
  \caption{The singular values of the Yukawa coupling matrices and the CKM mixing angles and CP phase in MSSM
  at $\mu=\mu_{\rm GUT}=2\cdot10^{16}$~GeV.
  Also shown are the errors of the quark Yukawa couplings, propagated from the experimental errors of the corresponding masses,
 and the maximal errors of the CKM parameters, obtained by assuming maximal correlation of the experimental errors of the Wolfenstein parameters.}
   \begin{tabular}{|c||c|} \hline
                                      & Value with Eq.~(\ref{massspectrum})  \\ \hline
    $y_u$           &2.81(11)$\times10^{-6}$          \\
    $y_c$           &0.001433(16)                               \\ 
    $y_t$            &0.4722(58)                                    \\ \hline
    $y_d$           &0.0003141(33)                             \\
    $y_s$           &0.006243(38)                                 \\ 
    $y_b$           &0.3557(16)                                   \\ \hline
    $y_e$           &0.0001261                                 \\
    $y_\mu$           &0.02662                                \\
    $y_\tau$           &0.5095                                  \\ \hline
    $\cos\theta_{13}^{\rm ckm}\sin\theta_{12}^{\rm ckm}$            & 0.22500(24)      \\
    $\cos\theta_{13}^{\rm ckm}\sin\theta_{23}^{\rm ckm}$           & 0.04171(70)         \\
    $\sin\theta_{13}^{\rm ckm}$                                                              & 0.00367(20)      \\
    $\delta_{\rm km}$~(rad)                                                                       &1.148(33)            \\ \hline
  \end{tabular}
  \label{values}
  \end{center}
\end{table}
\noindent
Also, we evaluate one-loop RG corrections to the Wilson coefficient of the Weinberg operator.
We write the Weinberg operator in MSSM as Eq.~(\ref{weinberg}) and that in SM as $-{\cal L}=\frac{1}{2}(C'_\nu)_{ij}\,\psi_{L_i}\psi_{L_j}H H$
 where $\psi_{L_i}$ denote the lepton doublets and $H$ the Higgs field.
We express the one-loop RG corrections in MSSM and SM as $C_\nu(\mu)=R(\mu)C_\nu(\mu_{\rm SUSY})R^T(\mu)$, $C'_\nu(\mu)=R'(\mu)C'_\nu(M_Z)R'^T(\mu)$, respectively,
 and perform the matching as $C_\nu(\mu_{\rm SUSY})=C'_\nu(\mu_{\rm SUSY})$, since $\tan\beta\gg1$.
We solve the one-loop RG equations and calculate the product of $R(\mu_{\rm GUT})$ and $R'(\mu_{\rm SUSY})$.
Here we approximate the scale of the Majorana mass $(M_N)_{ij}$ to be $\mu_{\rm GUT}$.
The product of $R(\mu_{\rm GUT})$ and $R'(\mu_{\rm SUSY})$ in the flavor basis where the lepton doublets have a diagonal Yukawa coupling matrix, is calculated as
\bea
R(\mu_{\rm GUT})R'(\mu_{\rm SUSY}) =    \begin{pmatrix} 
      1.09 & 0 & 0 \\
      0 & 1.09 & 0 \\
      0 & 0 & 1.14 \\
   \end{pmatrix}.
 \label{rr}
\eea
\\

We fit the quark and charged lepton Yukawa couplings and the CKM parameters at $\mu=\mu_{\rm GUT}$ in Table~\ref{values}
 with the Yukawa coupling matrices $Y_{10},Y_{126},Y_{120}$ and numbers $r_1,r_2,r_3,r_e$ based on Eqs.~(\ref{yu})-(\ref{ye}).
Also, we calculate the neutrino mass matrix up to the overall constant, which is proportional to $C'_\nu(M_Z)$,
 from $Y_{10},Y_{126},Y_{120}$ and $r_2,r_D$ using Eqs.~(\ref{ydirac})-(\ref{cnu}),(\ref{rr}), and with it we fit the neutrino oscillation data in NuFIT~5.1~\cite{Esteban:2020cvm,nufit}.
Meanwhile, we minimize Eq.~(\ref{tobeminimized}).

We restrict the parameter space to the region with $r_3=0$, since it is easier to minimize Eq.~(\ref{tobeminimized}) when $r_3=0$.
This is because when $r_3=0$, $Y_u=Y_{10}+r_2Y_{126}$ holds and we get
 $|(Y_{10})_{u_Li}+r_2(Y_{126})_{u_Li}|=|(Y_u)_{u_L i}|\leq y_u$ and $|(Y_{10})_{u_Ri}+r_2(Y_{126})_{u_Ri}|=|(Y_u)_{u_R i}|\leq y_u$ for any flavor index $i$.
Under the restriction of $r_3=0$, we parametrize the Yukawa coupling matrices as follows:
We go to the flavor basis where the isospin-doublet down-type quark components of ${\bf 16}^i$ matter fields have a diagonal Yukawa coupling matrix.
Since $Y_u=Y_{10}+r_2Y_{126}$ is symmetric, $Y_u$ in this basis can be written as
\bea
Y_u=V_{\rm CKM}^T   
\begin{pmatrix}
      y_u & 0 & 0 \\
      0 & y_c \,e^{2i \,d_2} & 0 \\
      0 & 0 & y_t \,e^{2i \,d_3}\\
   \end{pmatrix}
   V_{\rm CKM},
   \label{yu3}
\eea
 where $V_{\rm CKM}$ denotes the CKM matrix and $d_2,d_3$ are undetermined phases.
In the same flavor basis, $Y_d$ can be written as
\bea
Y_d=\begin{pmatrix} 
      y_d & 0 & 0 \\
      0 & y_s & 0 \\
      0 & 0 & y_b \\
   \end{pmatrix}
   V_{dR},
   \label{yd3}
 \eea
 where $V_{dR}$ is an undetermined unitary matrix.
From Eqs.~(\ref{yu})-(\ref{ydirac}), $Y_e$ and $Y_D$ are written in terms of $Y_u,Y_d$ as
\bea
&&\frac{1}{r_1}Y_e=Y_u-(3+r_2)Y_{126}+r_e\frac{1}{r_1}\frac{1}{2}\left(Y_d-Y_d^T\right),
\\
&&Y_D=Y_u-4r_2Y_{126}+r_D\frac{1}{r_1}\frac{1}{2}\left(Y_d-Y_d^T\right)
\eea
 with
 \bea
 Y_{126} = \frac{1}{1-r_2}\left\{\frac{1}{r_1}\frac{1}{2}\left(Y_d+Y_d^T\right)-Y_u\right\}.
 \eea
The Majorana mass Eq.~(\ref{majorana}) is found to satisfy
\bea
M_N \propto Y_{126}.
\eea

The analysis of fitting and minimization proceeds with the above parameterization as follows:
We fix $y_u,y_c,y_t$ and the parameters of the CKM matrix at the central values in Table~\ref{values}.
Then we randomly generate $y_d/r_1,\,y_s/r_1,\,y_b/r_1$, phases $d_2,d_3$, unitary matrix $V_{dR}$,
 and complex numbers $r_2,r_e,r_D$, and calculate the singular values of $\frac{1}{r_1}Y_e$.
We determine $r_1$ by requiring that the smallest singular value of $Y_e$ equal the value of $y_e$ in Table~\ref{values}.
Then we require that the values of $y_d,y_s,y_b$ be within the 3$\sigma$ ranges in Table~\ref{values}, and the first and second largest singular values of $Y_e$
 respectively be within $\pm$0.1\% ranges of the values of $y_\tau,y_\mu$
 \footnote{Since the experimental errors of the charged lepton Yukawa couplings are negligibly small,
  here we loosen the fitting criteria.}.
Because the active neutrino mass matrix $M_\nu$ is proportional to $C'_\nu(M_Z)$, we calculate $M_\nu$ up to the overall constant as 
\bea
M_\nu \ \propto \ \left(R(\mu_{\rm GUT})R'(\mu_{\rm SUSY})\right)^{-1} Y_D Y_{126}^{-1} Y_D^T \left(R(\mu_{\rm GUT})R'(\mu_{\rm SUSY})\right)^{T\,-1}.
\label{neutrinomassmatrix}
\eea
Then we calculate from Eq.~(\ref{neutrinomassmatrix}) the three neutrino mixing angles 
 $\sin^2\theta_{12}$, $\sin^2\theta_{23}$, $\sin^2\theta_{13}$ and the ratio of the neutrino mass squared differences $\Delta m_{21}^2/\Delta m_{31}^2$,
 and require them to be within the 3$\sigma$ ranges of NuFIT5.1 results (with SK atmospheric data).
Here we assume the normal mass hierarchy, because it is almost impossible to realize the inverted mass hierarchy from $Y_D,Y_{126}$ as these matrices
 have hierarchical structures.
Finally, we select sets of values of $y_d/r_1,\,y_s/r_1,\,y_b/r_1$, $d_2,d_3$, $V_{dR}$, $r_2,r_e,r_D$ that meet the above fitting criteria,
 calculate Eq.~(\ref{tobeminimized}) from the sets, and minimize the value of Eq.~(\ref{tobeminimized}).
\\

\subsection{Result}
\label{section-fitting2}

From the analysis of Section~\ref{section-fitting1}, we have obtained the values of the Yukawa coupling matrices $Y_{10},Y_{126},Y_{120}$ and numbers $r_1,r_2,r_e,r_D$ in Appendix~B. There, components of $Y_{10},Y_{126}$ satisfy
\begin{align}
&|(Y_{10})_{u_R d_R}|/y_u=1.4, \ \ \ |(Y_{126})_{u_R d_R}|/y_u=1.9, \ \ \ |(Y_{10})_{u_R s_R}|/y_u=2.0, \ \ \ |(Y_{126})_{u_R s_R}|/y_u=2.0,
\nn\\
&|(Y_{10})_{u_L d_L}|/y_u=1.9, \ \ \ |(Y_{126})_{u_L d_L}|/y_u=1.8, \ \ \ |(Y_{10})_{u_L s_L}|/y_u=0.33, \ \ \ |(Y_{126})_{u_L s_L}|/y_u=0.34,
\nn\\
&|(Y_{10})_{u_L u_L}|/y_u=0.45, \ \ \ |(Y_{126})_{u_L u_L}|/y_u=0.87.
\end{align}
Clearly, the conditions that the components of Eq.~(\ref{tobeyu}) be on the order of the up quark Yukawa coupling are satisfied.

\begin{figure}[H]
\begin{center}
\includegraphics[width=80mm]{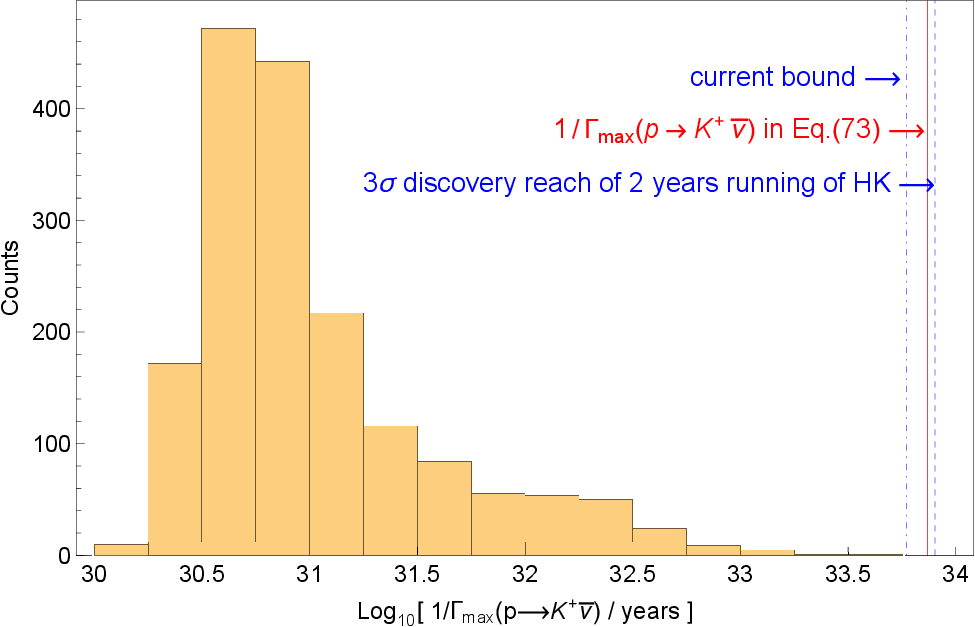}
\includegraphics[width=80mm]{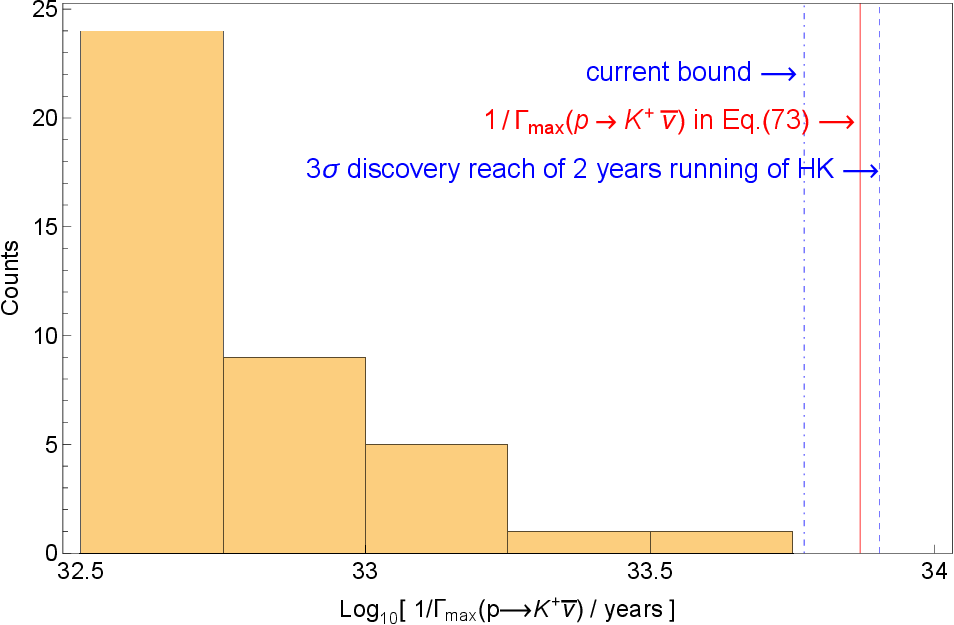}
\\
\includegraphics[width=80mm]{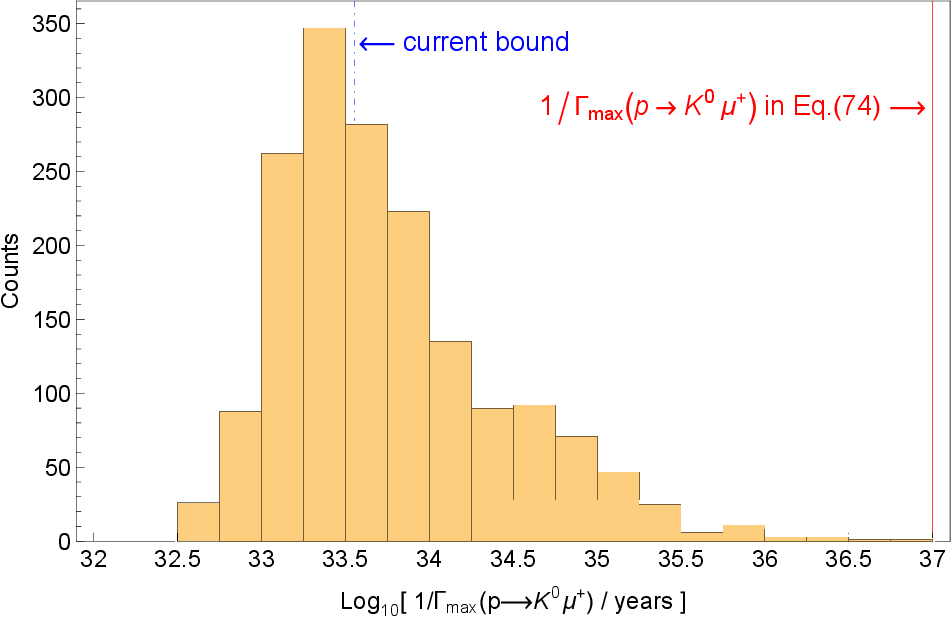}
\includegraphics[width=80mm]{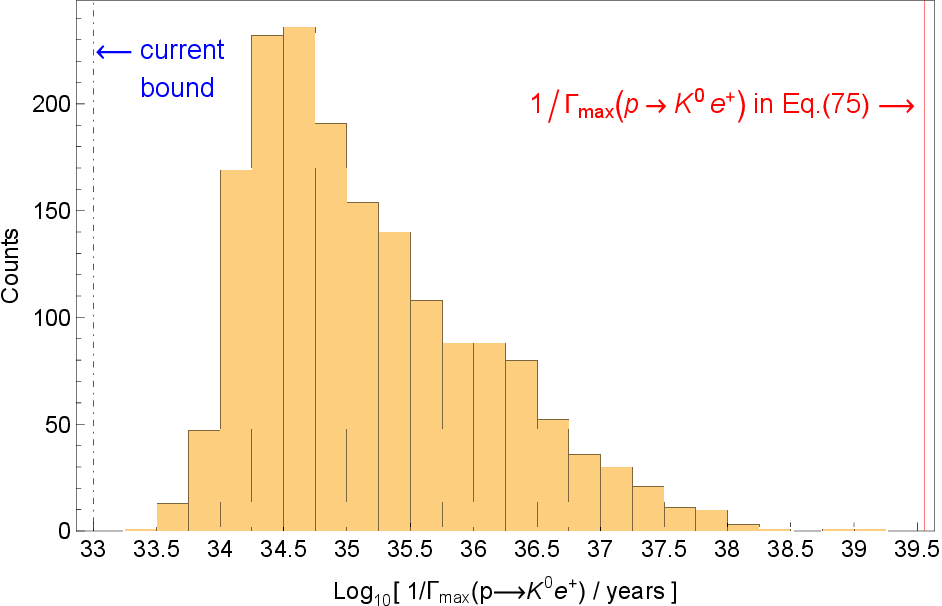}
\caption{
Distributions of the ``minimal proton partial lifetimes" calculated from multiple results of ``fitting without minimizing Eq.~(\ref{tobeminimized})".
The upper two panels show the distribution of $1/\Gamma_{\rm max}(p\to K^+ \bar{\nu})$, where the upper-right panel magnifies the right tail of the upper-left one.
The blue dot-dashed line indicates the current bound on the $p\to K^+ \bar{\nu}$ partial lifetime~\cite{Super-Kamiokande:2014otb}, the red solid line the value of $1/\Gamma_{\rm max}(p\to K^+ \bar{\nu})$ calculated from the Yukawa coupling matrices of Appendix~B in Eq.~(\ref{knunum}), and the blue dashed line the $3\sigma$ discovery reach of {\bf 2 years} running of Hyper-Kamiokande~\cite{Hyper-Kamiokande:2018ofw}.
The lower-left panel shows the distribution of
$1/\Gamma_{\rm max}(p\to K^0 \mu^+)$, where the blue dot-dashed line indicates the current bound on the $p\to K^0 \mu^+$ partial lifetime~\cite{Super-Kamiokande:2022egr} (part of the line is hidden behind the histogram), 
 and the red solid line the value of $1/\Gamma_{\rm max}(p\to K^0 \mu^+)$ calculated from the Yukawa coupling matrices of Appendix~B in Eq.~(\ref{kmunum}).
The lower-right panel shows the distribution of $1/\Gamma_{\rm max}(p\to K^0 e^+)$, where the blue dot-dashed line indicates the current bound on the $p\to K^0 e^+$ partial lifetime~\cite{Super-Kamiokande:2005lev}, 
 and the red solid line the value of $1/\Gamma_{\rm max}(p\to K^0 e^+)$ calculated from the Yukawa coupling matrices of Appendix~B in Eq.~(\ref{kenum}).
}
\label{histogram}
\end{center}
\end{figure}

We evaluate how dimension-five proton decay is suppressed when the conditions are met.
To this end, we compare ``minimal proton partial lifetimes" calculated from the Yukawa coupling matrices of Appendix~B,
 with those calculated from results of ``fitting without minimizing Eq.~(\ref{tobeminimized})" where we only
 fit the quark and charged lepton Yukawa couplings, CKM parameters and neutrino oscillation data as Section~\ref{section-fitting1} but do not minimize Eq.~(\ref{tobeminimized}) so that the conditions are not necessarily satisfied.
Here the ``minimal proton partial lifetimes", $1/\Gamma_{\rm max}(p\to K^+ \bar{\nu})$, $1/\Gamma_{\rm max}(p\to K^0 \mu^+)$, $1/\Gamma_{\rm max}(p\to K^0 e^+)$,
 are defined as ($\beta=e,\mu$)
\begin{align}
&\Gamma_{\rm max}(p\to K^+ \bar{\nu}) = \frac{m_N}{64\pi}\left(1-\frac{m_K^2}{m_N^2}\right)^2
\nn\\
&\ \ \ \ \ \ \ \ \ \ \ \ \ \ \ \times\left(|A_{\rm max}(p\to K^+ \bar{\nu}_\tau)|^2+|A_{\rm max}(p\to K^+ \bar{\nu}_\mu)|^2+|A_{\rm max}(p\to K^+ \bar{\nu}_e)|^2\right),
\label{minimalprotonpartiallifetime} \\
&\Gamma_{\rm max}(p\to K^0 \beta^+) = \frac{m_N}{64\pi}\left(1-\frac{m_K^2}{m_N^2}\right)^2|A_{\rm max}(p\to K^0 \beta^+)|^2,
\label{minimalprotonpartiallifetime2}
\end{align}
 where
\begin{align}
A_{\rm max}(p\to K^+ \bar{\nu}_\tau) &= |A_{\rm max}(p\to K^+ \bar{\nu}_\tau)_{{\rm from} \ C_{5R}}|+|A_{\rm max}(p\to K^+ \bar{\nu}_\tau)_{{\rm from} \ C_{5L}}|
\label{ataufromc5randc5l}
\end{align}
 and
\begin{align}
&A_{\rm max}(p\to K^+ \bar{\nu}_\tau)_{{\rm from} \ C_{5R}}
=
\alpha_H(\mu_{\rm had})\frac{1}{f_\pi}A_{RL}(\mu_{\rm had},\mu_{\rm SUSY}) \frac{\mu_H}{m_{\tilde{t}_R}^2}
\frac{1}{16\pi^2}{\cal F}'\, y_t y_\tau\,
A_R^{\tau t}(\mu_{\rm SUSY},\mu_{\rm GUT})\frac{1}{M_{H_C}}
\nn\\
&\ \ \ \ \ \ \ \ \times \max_{A,B}\left\{ \left| 
\left(1+\frac{D}{3}+F\right)(V^{\rm ckm}_{ts})^* \left((Y_{A})_{\tau_R t_R}(Y_{B})_{u_Rd_R}-(Y_{A})_{\tau_R u_R}(Y_{B})_{t_Rd_R}\right) \right.\right.
\nn\\
&\left.\left.\ \ \ \ \ \ \ \ \ \ \ \ \ \ \ \ \ \ \ \ \ \ \ \ \ \ \ \ \ \ \ \ \ \ \ \ + \frac{2D}{3}(V^{\rm ckm}_{td})^* \left((Y_{A})_{\tau_R t_R}(Y_{B})_{u_Rs_R}-(Y_{A})_{\tau_R u_R}(Y_{B})_{t_Rs_R}\right) \right| \right\},
\label{ataufromc5r}\\
&A_{\rm max}(p\to K^+ \bar{\nu}_\tau)_{{\rm from} \ C_{5L}}
=\beta_H(\mu_{\rm had})\frac{1}{f_\pi}A_{LL}(\mu_{\rm had},\mu_{\rm SUSY}) \frac{M_{\widetilde{W}}}{m_{\tilde{q}}^2}\frac{1}{16\pi^2}{\cal F} \, g_2^2 \, A_L^\tau(\mu_{\rm SUSY},\mu_{\rm GUT})\frac{1}{M_{H_C}}
\nn\\
&\ \ \ \ \ \ \times \max_{A',B'}\left\{ \left| 
\left(1+\frac{D}{3}+F\right)(Y_{A'})_{u_Ld_L}(Y_{B'})_{s_L\tau_L} + \frac{2D}{3}(Y_{A'})_{u_Ls_L}(Y_{B'})_{d_L\tau_L}\right| \right\},
\label{ataufromc5l}\\
&A_{\rm max}(p\to K^+ \bar{\nu}_\beta)
=\beta_H(\mu_{\rm had})\frac{1}{f_\pi}A_{LL}(\mu_{\rm had},\mu_{\rm SUSY}) \frac{M_{\widetilde{W}}}{m_{\tilde{q}}^2}\frac{1}{16\pi^2}{\cal F} \, g_2^2 \, A_L^\beta(\mu_{\rm SUSY},\mu_{\rm GUT})\frac{1}{M_{H_C}}
\nn\\
&\ \ \ \ \ \ \times \max_{A',B'}\left\{ \left| 
\left(1+\frac{D}{3}+F\right)(Y_{A'})_{u_Ld_L}(Y_{B'})_{s_L\beta_L} + \frac{2D}{3}(Y_{A'})_{u_Ls_L}(Y_{B'})_{d_L\beta_L}\right| \right\},
\label{abeta}\\
&A_{\rm max}(p\to K^0 \beta^+)=\beta_H(\mu_{\rm had})\frac{1}{f_\pi}A_{LL}(\mu_{\rm had},\mu_{\rm SUSY}) \frac{M_{\widetilde{W}}}{m_{\tilde{q}}^2}\frac{1}{16\pi^2}{\cal F} \, g_2^2 \, A_L^\beta(\mu_{\rm SUSY},\mu_{\rm GUT})\frac{1}{M_{H_C}}
\nn\\
&\ \ \ \ \ \ \times (1-D+F) \ \max_{A',B'}\left\{ \left| (Y_{A'})_{u_Ls_L}(Y_{B'})_{u_L\beta_L}-(Y_{A'})_{u_Lu_L}(Y_{B'})_{s_L\beta_L} \right| \right\},
\label{abetaprime}
\end{align}
 where $A,B$ in Eq.~(\ref{ataufromc5r}) run as $(A,B)=(10,10)$, (10,126), (126,10), (126,126), (120,10), (120,126), (120,120),
 and $A',B'$ in Eqs.~(\ref{ataufromc5l})-(\ref{abetaprime}) run as $(A',B')=(10,10)$, (10,126), (126,10), (126,126).
We assume the SUSY particle spectrum of Eq.~(\ref{massspectrum}) and take $M_{H_C}=2\cdot10^{16}$~GeV in the calculation.
Note that Eqs.~(\ref{ataufromc5r})-(\ref{abetaprime}) take into account the texture of the colored Higgs mass matrix satisfying Eqs.~(\ref{coloredhiggstex1}),(\ref{coloredhiggstex2}).
An implication of Eqs.~(\ref{ataufromc5randc5l})-(\ref{abetaprime}) is that we estimate the maximal values of the amplitudes without specifying the $O(1)$ numbers $a,b,d,e,g,h,j$ in Eqs.~(\ref{yyr1})-(\ref{yyl3}) and the relative phase between the Higgsino mass and Wino mass.
The ``minimal proton partial lifetimes" calculated from the Yukawa coupling matrices of Appendix~B are
\footnote{
These values are consistent with the current experimental bounds on the $p\to K^+ \bar{\nu}$ partial lifetime~\cite{Super-Kamiokande:2014otb}
 and on the $p\to K^0 \mu^+/e^+$ partial lifetimes~\cite{Super-Kamiokande:2022egr,Super-Kamiokande:2005lev},
 which justifies our choice of the benchmark SUSY particle mass spectrum~Eq.~(\ref{massspectrum}).
}
\bea
1/\Gamma_{\rm max}(p\to K^+ \bar{\nu}) = 7.4\times10^{33}~{\rm years},
\label{knunum}\\
1/\Gamma_{\rm max}(p\to K^0 \mu^+) = 1.0\times10^{37}~{\rm years},
\label{kmunum}\\
1/\Gamma_{\rm max}(p\to K^0 e^+) = 3.6\times10^{39}~{\rm years}.
\label{kenum}
\eea
On the other hand, the ``minimal proton partial lifetimes" calculated from multiple results of ``fitting without minimizing Eq.~(\ref{tobeminimized})" are distributed as Fig.~\ref{histogram}.

\noindent
In Fig.~\ref{histogram}, we overlay the values calculated from the Yukawa coupling matrices of Appendix~B in Eqs.~(\ref{knunum})-(\ref{kenum}). Also, the current bounds on the $p\to K^+ \bar{\nu}$~\cite{Super-Kamiokande:2014otb}, $p\to K^0 \mu^+$~\cite{Super-Kamiokande:2022egr}, $p\to K^0 e^+$~\cite{Super-Kamiokande:2005lev} partial lifetimes, and the $3\sigma$ discovery reach of 2 years running of Hyper-Kamiokande~\cite{Hyper-Kamiokande:2018ofw} are shown.

Comparing Eqs.~(\ref{knunum})-(\ref{kenum}) with Fig.~\ref{histogram}, we see that the proton partial lifetimes calculated from the Yukawa coupling matrices of Appendix~B are on the upper edge of the distributions of proton partial lifetimes calculated from results of ``fitting without minimizing Eq.~(\ref{tobeminimized})".
This confirms that the texture of the Yukawa coupling matrices satisfying the conditions that {\it the components of Eq.~(\ref{tobeyu}) be on the order of the up quark Yukawa coupling}, contributes to suppressing dimension-five proton decay.
Specifically, the benchmark SUSY particle mass spectrum Eq.~(\ref{massspectrum}), where the SUSY particle masses are all at 1500~TeV and $\tan\beta=50$, is consistent with all the experimental bounds on proton partial lifetimes if the above conditions are satisfied.
On the other hand, if these conditions are not met, this benchmark almost always violates the bound on the $p\to K^+ \bar{\nu}$ partial lifetime.
We also see that for this benchmark mass spectrum, when the above conditions are satisfied,
 we expect to discover the $p\to K^+ \bar{\nu}$ decay with 2 years running of Hyper-Kamiokande.
\\

In addition to the texture of the Yukawa coupling matrices, we have required that the colored Higgs mass matrix be tuned such that Eqs.~(\ref{coloredhiggstex1}),(\ref{coloredhiggstex2}) hold.
Now we examine the degree of tuning of the colored Higgs mass matrix necessary to suppress dimension-five proton decay.
To this end, we consider non-zero $c,f$ and deviations of $a/d$ and $b/e$ from the relations of Eq.~(\ref{coloredhiggstex2}),
 and evaluate maximum values of $|c|,|f|$ and maximum deviations of $a/d$ and $b/e$ that reduce
 $1/\Gamma_{\rm max}(p\to K^+ \bar{\nu})$ from Eq.~(\ref{knunum}) by at most 20\% (with the same Yukawa coupling matrices).
Here the phases of $c,f,a/d,b/e$ are chosen such that they reduce $1/\Gamma_{\rm max}(p\to K^+ \bar{\nu})$ maximally,
 and the contributions of $c$, $f$, $a/d$, $b/e$ are studied separately.
We numerically find that the maximum values of $|c|,|f|$ are
\bea
|c| = 0.14, \ \ \ \ \ \ \ \ |f| = 0.14,
\label{cfmax}
\eea
 and the maximum deviations of $a/d$ and $b/e$ are
\bea
\left|a + d\frac{(Y_{126})_{d_Ls_L}}{(Y_{10})_{d_Ls_L}}\right| = 0.0097,
\ \ \ \ \ \ \ \ \left|b + e\frac{(Y_{126})_{d_Ls_L}}{(Y_{10})_{d_Ls_L}}\right| = 0.011.
\eea
Interestingly, the requirement of $c=f=0$ is not so severe, 
 while the conditions of $a\, (Y_{10})_{d_L s_L}+d\, (Y_{126})_{d_L s_L}=0$ and $b\, (Y_{10})_{d_L s_L}+e\, (Y_{126})_{d_L s_L}=0$
 must be satisfied with 1\% precision.
For the other decay modes, the deviations of $a/d$ and $b/e$ do not affect $1/\Gamma_{\rm max}(p\to K^0 \mu^+)$ and $1/\Gamma_{\rm max}(p\to K^0 e^+)$.
Non-zero $c,f$ whose absolute values are below Eq.~(\ref{cfmax}) do not alter $1/\Gamma_{\rm max}(p\to K^0 \mu^+)$ and $1/\Gamma_{\rm max}(p\to K^0 e^+)$ because the products of Yukawa coupling components associated with $c$ or $f$ in Eq.~(\ref{yyl3}) are numerically smaller than 0.14 times the largest product of Yukawa coupling components in Eq.~(\ref{yyl3}).
\\

We comment that the Yukawa coupling matrices and coefficients in Appendix~B
 give a prediction on poorly or not measured neutrino parameters, which are the Dirac CP phase of the neutrino mixing matrix, the sum of the neutrino mass, and the effective neutrino mass for neutrinoless double $\beta$ decay.
The prediction is shown in Appendix~C.

We comment on other nucleon decay modes.
The $N\to \pi\beta^+$ and $p\to \eta \beta^+$ decays are subdominant compared to the $p\to K^0 \beta^+$ decays, because the amplitudes
 of $N\to \pi\beta^+$ and $p\to \eta \beta^+$ involve the same Yukawa coupling components as those of $p\to K^0 \beta^+$ except that $s_L$ is replaced by $d_L$.
Nevertheless, observation of $N\to \pi\beta^+$ and $p\to \eta \beta^+$ along with $p\to K^0 \beta^+$
 may provide an experimental clue to the texture of the Yukawa coupling matrices. Hence, we present in Appendix~D
 the ``minimal partial lifetimes" of these modes calculated from the Yukawa coupling matrices of Appendix~B through the formulas in Ref.~\cite{Nath:2006ut}.
\\

\section{Summary}
\label{summary}

We have pursued the possibility that dimension-five proton decay is suppressed by a texture of the Yukawa coupling matrices
 in the general renormalizable SUSY $SO(10)$ GUT model where Yukawa coupling matrices of ${\bf 16}$ representation matter fields with ${\bf 10},{\bf \overline{126}},{\bf 120}$ fields $Y_{10},Y_{126},Y_{120}$ give the quark and lepton Yukawa couplings and Majorana mass of the singlet neutrinos.
We have derived conditions for a texture of the Yukawa coupling matrices suppressing dimension-five proton decay, which state that
 {\it components
 $(Y_{10})_{u_R d_R}$, $(Y_{126})_{u_R d_R}$,
 $(Y_{10})_{u_R s_R}$, $(Y_{126})_{u_R s_R}$, $(Y_{10})_{u_L d_L}$, $(Y_{126})_{u_L d_L}$, $(Y_{10})_{u_L u_L}$, 
 $(Y_{126})_{u_L u_L}$, $(Y_{10})_{u_L s_L}$, $(Y_{126})_{u_L s_L}$
 should be all on the order of the up quark Yukawa coupling} $y_u$.
Additionally, the colored Higgs mass matrix should satisfy Eqs.~(\ref{coloredhiggstex1}),(\ref{coloredhiggstex2}).
We have obtained the values of the Yukawa coupling matrices that satisfy the above conditions and that are consistent with
 the experimental data of quark and lepton masses and flavor mixings.
By comparing the ``minimal proton partial lifetimes" calculated from the Yukawa coupling matrices that meet the conditions and those that do not necessarily so, we have confirmed that the texture of the Yukawa coupling matrices satisfying the conditions contributes to suppressing dimension-five proton decay.
Specifically, we have found that a SUSY particle mass spectrum where the SUSY particle masses are all at 1500~TeV and $\tan\beta=50$
 is consistent with all the experimental bounds on proton decay if the above conditions are satisfied.
Also, for this mass spectrum, when the conditions are met, we expect to discover the $p\to K^+ \bar{\nu}$ decay with 2 years running of Hyper-Kamiokande.
\\

\section*{Acknowledgement}
This work is partially supported by Scientific Grants by the Ministry of Education, Culture, Sports, Science and Technology of Japan,
No.~21H00076 (NH) and No.~19K147101 (TY).
\\

\section*{Appendix~A: Superpotential}

We review our definition of the coupling constants and masses for
 $H$, $\Delta$, $\overline{\Delta}$, $\Sigma$, $\Phi$, $A$ fields in 
 ${\bf 10}$, ${\bf 126}$, ${\bf \overline{126}}$, ${\bf 120}$, ${\bf 210}$, ${\bf 45}$ representations,
 which follows Eq.~(2) of Ref.~\cite{Fukuyama:2004ps}.
The couplings are defined in the same way as Eq.~(3) of Ref.~\cite{Fukuyama:2004ps}.
Note that ${\bf 120}$ representation field is written as $D$ in Ref.~\cite{Fukuyama:2004ps}, while we write it as $\Sigma$.
The coupling constants are defined as
\bea
W &=& \frac{1}{2}m_1\Phi^2 + m_2 \overline{\Delta}\Delta + \frac{1}{2}m_3H^2
\nn\\ &+& \frac{1}{2}m_4A^2 + \frac{1}{2}m_6\Sigma^2
\nn\\ &+& \lambda_1\Phi^3 + \lambda_2\Phi\overline{\Delta}\Delta + (\lambda_3\Delta+\lambda_4\overline{\Delta})H\Phi
\nn\\ &+& \lambda_5 A^2\Phi - i\lambda_6A\overline{\Delta}\Delta + \frac{\lambda_7}{120}\varepsilon A\Phi^2
\nn\\ &+& \lambda_{15}\Sigma^2\Phi
\nn\\ &+& \Sigma\{\lambda_{16}HA + \lambda_{17}H\Phi + (\lambda_{18}\Delta+\lambda_{19}\overline{\Delta})A + (\lambda_{20}\Delta+\lambda_{21}\overline{\Delta})\Phi\}
\label{gutsuperpotential}
\eea
 where $\varepsilon$ denotes the antisymmetric tensor in $SO(10)$ space.
\\

\section*{Appendix~B: Values of $Y_{10}$, $Y_{126}$, $Y_{120}$ and $r_1,r_2,r_e,r_D$}

We present the values of the Yukawa coupling matrices $Y_{10}$, $Y_{126}$, $Y_{120}$ and numbers $r_1,r_2,r_e,r_D$ obtained from the analysis of Section~\ref{section-fitting1}.
 $Y_{10}$, $Y_{126}$, $Y_{120}$ are shown repeatedly in three different flavor bases.
For reference, the central value of the up quark Yukawa coupling at scale $\mu=\mu_{\rm GUT}=2\cdot10^{16}$~GeV in $\ov{{\rm DR}}$ scheme is $y_u=2.81\cdot10^{-6}$.
\begin{align}
   \begin{pmatrix} 
      (Y_{10})_{u_Rd_R} & (Y_{10})_{u_Rs_R} & (Y_{10})_{u_Rb_R}\\
      (Y_{10})_{c_Rd_R} & (Y_{10})_{c_Rs_R} & (Y_{10})_{c_Rb_R}\\
      (Y_{10})_{t_Rd_R} & (Y_{10})_{t_Rs_R} & (Y_{10})_{t_Rb_R}\\
   \end{pmatrix}&=\begin{pmatrix} 
      3.92\cdot10^{-6}\,e^{2.26\,i} &  5.61\cdot10^{-6}\,e^{-1.26\,i} & 0.00168\,e^{-0.21\,i}  \\
      0.00187\,e^{-1.22\,i}             &  0.00810\,e^{-1.24\,i}              & 0.0194\,e^{-1.04\,i} \\
      0.00592\,e^{-2.61\,i}             &  0.0202\,e^{-2.81\,i}                & 0.230\,e^{0.46\,i}     \\
    \end{pmatrix}     
\\  
   \begin{pmatrix} 
      (Y_{126})_{u_Rd_R} & (Y_{126})_{u_Rs_R} & (Y_{126})_{u_Rb_R}\\
      (Y_{126})_{c_Rd_R} & (Y_{126})_{c_Rs_R} & (Y_{126})_{c_Rb_R}\\
      (Y_{126})_{t_Rd_R} & (Y_{126})_{t_Rs_R} & (Y_{126})_{t_Rb_R}\\
   \end{pmatrix}&=\begin{pmatrix} 
     5.23\cdot10^{-6}\,e^{-2.28\,i} & 5.70\cdot10^{-6}\,e^{0.89\,i} & 0.00158\,e^{2.01\,i}    \\
     0.00198\,e^{1.12\,i}               & 0.00856\,e^{1.10\,i}              & 0.0182\,e^{1.18\,i}   \\
     0.00618\,e^{0.57\,i}               & 0.0197\,e^{0.31\,i}                & 0.228\,e^{-0.46\,i}    \\
    \end{pmatrix}
 \\  
   \begin{pmatrix} 
      (Y_{120})_{u_Rd_R} & (Y_{120})_{u_Rs_R} & (Y_{120})_{u_Rb_R}\\
      (Y_{120})_{c_Rd_R} & (Y_{120})_{c_Rs_R} & (Y_{120})_{c_Rb_R}\\
      (Y_{120})_{t_Rd_R} & (Y_{120})_{t_Rs_R} & (Y_{120})_{t_Rb_R}\\
   \end{pmatrix}&=\begin{pmatrix} 
      0.000365             & 0.00159                & 0.000457\,e^{2.44\,i}  \\
      0.00159\,e^{-3.13\,i}    & 0.000366\,e^{0.06\,i} & 0.000454\,e^{-0.26\,i}  \\
      0.000328\,e^{-1.94\,i}  & 0.000539\,e^{1.55\,i} & 1.71\cdot10^{-5}\,e^{2.95\,i}    \\
    \end{pmatrix}
\end{align}
\begin{align}
   \begin{pmatrix} 
      (Y_{10})_{u_Ld_L} & (Y_{10})_{u_Ls_L} & (Y_{10})_{u_Lb_L}\\
      (Y_{10})_{c_Ld_L} & (Y_{10})_{c_Ls_L} & (Y_{10})_{c_Lb_L}\\
      (Y_{10})_{t_Ld_L} & (Y_{10})_{t_Ls_L} & (Y_{10})_{t_Lb_L}\\
   \end{pmatrix}&=\begin{pmatrix} 
      5.45\cdot10^{-6}\,e^{-1.30\,i} & 0.920\cdot10^{-6}\,e^{1.39\,i} & 0.00168\,e^{0.58\,i}    \\
      0.00187\,e^{-2.28\,i}             & 0.00810\,e^{0.85\,i}            & 0.0194\,e^{-0.25\,i}   \\
      0.00385\,e^{1.98\,i}            & 0.0208\,e^{-0.72\,i}             & 0.230\,e^{1.25\,i}     \\
    \end{pmatrix}     
\\  
   \begin{pmatrix} 
      (Y_{126})_{u_Ld_L} & (Y_{126})_{u_Ls_L} & (Y_{126})_{u_Lb_L}\\
      (Y_{126})_{c_Ld_L} & (Y_{126})_{c_Ls_L} & (Y_{126})_{c_Lb_L}\\
      (Y_{126})_{t_Ld_L} & (Y_{126})_{t_Ls_L} & (Y_{126})_{t_Lb_L}\\
   \end{pmatrix}&=\begin{pmatrix} 
     5.10\cdot10^{-6}\,e^{0.41\,i} & 0.961\cdot10^{-6}\,e^{-2.01\,i} & 0.00158\,e^{2.80\,i}     \\
     0.00198\,e^{0.05\,i}             & 0.00856,e^{-3.10\,i}                & 0.0182\,e^{1.97\,i}   \\
     0.00395\,e^{-1.03\,i}              & 0.0206\,e^{2.44\,i}                  & 0.228\,e^{0.33\,i}    \\
    \end{pmatrix}
 \\  
   \begin{pmatrix} 
      (Y_{120})_{u_Ld_L} & (Y_{120})_{u_Ls_L} & (Y_{120})_{u_Lb_L}\\
      (Y_{120})_{c_Ld_L} & (Y_{120})_{c_Ls_L} & (Y_{120})_{c_Lb_L}\\
      (Y_{120})_{t_Ld_L} & (Y_{120})_{t_Ls_L} & (Y_{120})_{t_Lb_L}\\
   \end{pmatrix}&=\begin{pmatrix} 
       0.000365\,e^{-1.06\,i} & 0.00159\,e^{2.08\,i} & 0.000459\,e^{-3.06\,i}  \\
       0.00159\,e^{-1.05\,i}   & 0.000369\,e^{-1.10\,i} & 0.000452\,e^{0.54\,i}   \\
       0.000523\,e^{0.29\,i} & 0.000353\,e^{-2.51\,i} & 1.90\cdot10^{-5}\,e^{-2.51\,i}    \\
    \end{pmatrix} 
\end{align}
\begin{align}
   \begin{pmatrix} 
      (Y_{10})_{u_Lu_L} & (Y_{10})_{u_Lc_L} & (Y_{10})_{u_Lt_L}\\
       & (Y_{10})_{c_Lc_L} & (Y_{10})_{c_Lt_L}\\
       &  & (Y_{10})_{t_Lt_L}\\
   \end{pmatrix}&=\begin{pmatrix} 
       1.27\cdot10^{-6}\,e^{1.18\,i} & 7.11\cdot10^{-5}\,e^{0.60\,i} & 0.00168\,e^{0.58\,i} \\
                                                   &  0.00870\,e^{0.77\,i}     & 0.0192\,e^{-0.27\,i} \\
                                                   &                                     & 0.230\,e^{1.25\,i} \\
    \end{pmatrix}     
\label{y10ulul}\\  
   \begin{pmatrix} 
      (Y_{126})_{u_Lu_L} & (Y_{126})_{u_Lc_L} & (Y_{126})_{u_Lt_L}\\
       & (Y_{126})_{c_Lc_L} & (Y_{126})_{c_Lt_L}\\
       &  & (Y_{126})_{t_Lt_L}\\
   \end{pmatrix}&=\begin{pmatrix} 
       2.45\cdot10^{-6}\,e^{-1.39\,i} & 6.69\cdot10^{-5}\,e^{2.83\,i} & 0.00158\,e^{2.80\,i} \\
                                                  &  0.00907\,e^{3.11\,i}    & 0.0181\,e^{1.96\,i} \\
                                                  &                                    & 0.228\,e^{0.32\,i}  \\
    \end{pmatrix}
\label{y126ulul}\\  
   \begin{pmatrix} 
      (Y_{120})_{u_Lu_L} & (Y_{120})_{u_Lc_L} & (Y_{120})_{u_Lt_L}\\
       & (Y_{120})_{c_Lc_L} & (Y_{120})_{c_Lt_L}\\
       &  & (Y_{120})_{t_Lt_L}\\
   \end{pmatrix}&=\begin{pmatrix} 
         0  & 0.00164\,e^{2.09\,i} & 0.000435\,e^{-2.91\,i} \\
               &     0                       & 0.000457\,e^{0.54\,i}      \\
                 &                            &      0    \\
    \end{pmatrix} 
    \label{y120ulul}
\end{align}
\bea
r_1=0.871, \ \ \ r_2=1.06\,e^{0.92\,i}, \ \ \ r_e=1.01\,e^{0.79\,i}, \ \ \ r_D=0.482\,e^{-3.11\,i}.
\eea
In Eqs.~(\ref{y10ulul})-(\ref{y120ulul}), we do not display some off-diagonal components because in this flavor basis, $Y_{10},Y_{126}$ are symmetric and $Y_{120}$ is antisymmetric.
\\

\section*{Appendix~C: Prediction on Neutrino Parameters}

\begin{figure}[H]
\begin{center}
\includegraphics[width=80mm]{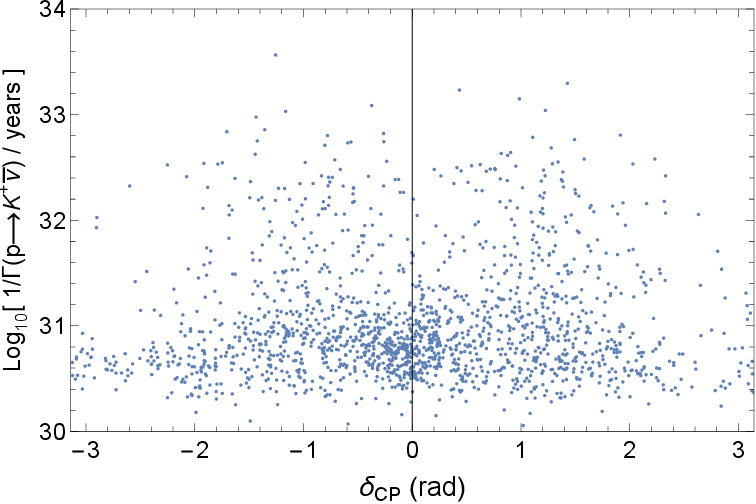}
\\
\includegraphics[width=80mm]{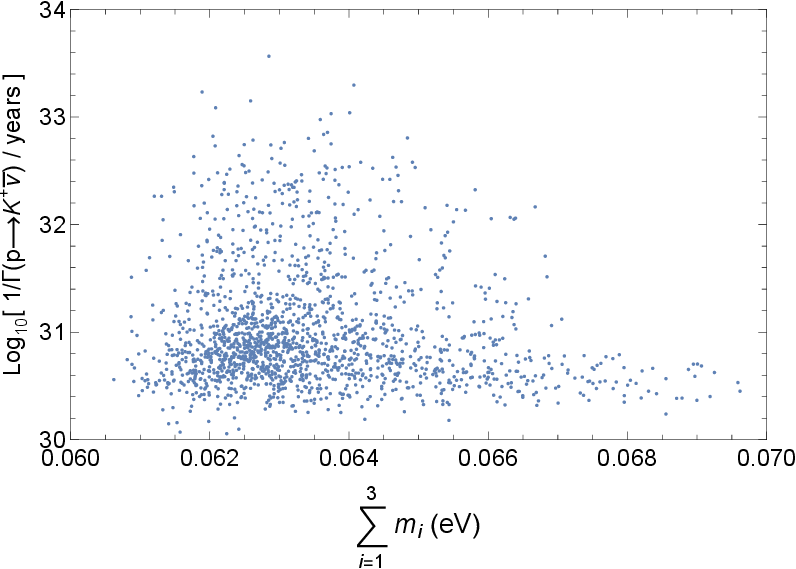}
\includegraphics[width=80mm]{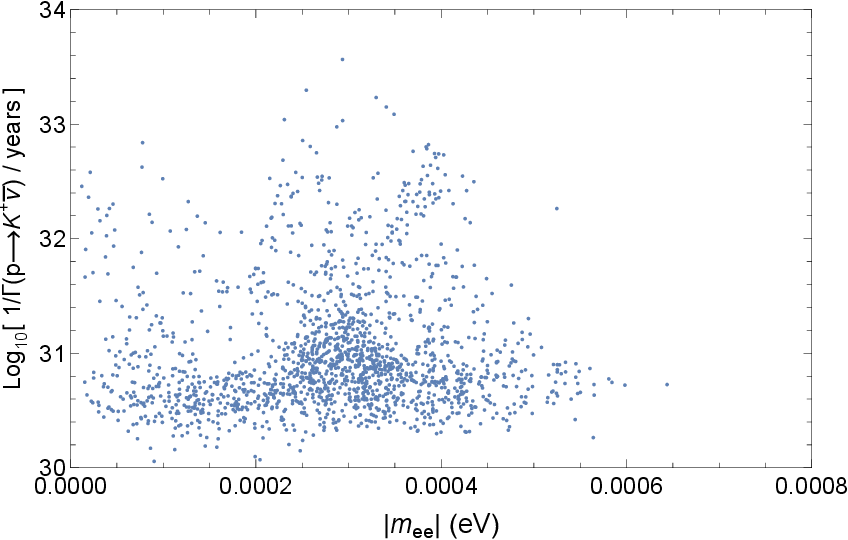}
\caption{
Results of ``fitting without minimizing Eq.~(\ref{tobeminimized})" on the planes of $1/\Gamma_{\rm max}(p\to K^+ \bar{\nu})$ versus $\delta_{\rm CP}$, $\sum_{i=1}^3 m_i$, $|m_{ee}|$ in the upper, lower-left and lower-right panels, respectively.
}
\label{neutrinopred}
\end{center}
\end{figure}

The result of the analysis of Section~\ref{section-fitting1}, shown in Appendix~B, gives the following prediction on
 the Dirac CP phase of the neutrino mixing matrix, $\delta_{\rm CP}$, the sum of the neutrino mass, $\sum_{i=1}^3 m_i$, and the effective neutrino mass for neutrinoless double $\beta$ decay, $|m_{ee}|$:
\begin{align}
\delta_{\rm CP}&=1.35~{\rm rad},
\label{prediction_delta}\\
\sum_{i=1}^3 m_i &= 0.0630~{\rm eV},
\\
|m_{ee}|&=0.000263~{\rm eV}.
\label{prediction_mee}
\end{align}
We caution that
 there is no clear correlation between the prediction on $\delta_{\rm CP}$, $\sum_{i=1}^3 m_i$, $|m_{ee}|$ and the degree of suppression of dimension-five proton decay, as seen in Fig.~\ref{neutrinopred} where we plot the results of ``fitting without minimizing Eq.~(\ref{tobeminimized})"
 on the planes of $1/\Gamma_{\rm max}(p\to K^+ \bar{\nu})$ versus $\delta_{\rm CP}$, $\sum_{i=1}^3 m_i$, $|m_{ee}|$.
Similar figures are obtained for $1/\Gamma_{\rm max}(p\to K^0 \mu^+)$ and $1/\Gamma_{\rm max}(p\to K^0 e^+)$.
Therefore, the prediction of Eqs.~(\ref{prediction_delta})-(\ref{prediction_mee}) is not a consequence of the texture of the Yukawa coupling matrices suppressing dimension-five proton decay.

\section*{Appendix~D: Other Nucleon Decay Modes}

The ``minimal partial lifetimes" of the $N\to \pi\beta^+$ and $p\to \eta \beta^+$ modes ($\beta=e,\mu$) defined analogously to Eq.~(\ref{minimalprotonpartiallifetime2}) and
 calculated from the Yukawa coupling matrices of Appendix~B through the formulas in Ref.~\cite{Nath:2006ut} are
\bea
1/\Gamma_{\rm max}(p\to \pi^0 \mu^+) = 1.5\times10^{37}~{\rm years},
\\
1/\Gamma_{\rm max}(p\to \pi^0 e^+) = 7.2\times10^{39}~{\rm years},
\\
1/\Gamma_{\rm max}(n\to \pi^- \mu^+) = 4.9\times10^{37}~{\rm years},
\\
1/\Gamma_{\rm max}(n\to \pi^- e^+) = 2.4\times10^{39}~{\rm years},
\\
1/\Gamma_{\rm max}(p\to \eta \mu^+) = 3.9\times10^{37}~{\rm years},
\\
1/\Gamma_{\rm max}(p\to \eta e^+) = 1.9\times10^{40}~{\rm years}.
\eea



\begin{thebibliography}{99}


\bibitem{Georgi:1974my} 
  H.~Georgi,
  ``The State of the Art—Gauge Theories,''
  AIP Conf.\ Proc.\  {\bf 23}, 575 (1975).
\bibitem{Fritzsch:1974nn} 
  H.~Fritzsch and P.~Minkowski,
  ``Unified Interactions of Leptons and Hadrons,''
  Annals Phys.\  {\bf 93}, 193 (1975).


\bibitem{Gell-Mann:1979vob}
M.~Gell-Mann, P.~Ramond and R.~Slansky,
``Complex Spinors and Unified Theories,''
Conf. Proc. C \textbf{790927}, 315-321 (1979)
[arXiv:1306.4669 [hep-th]].
\bibitem{Yanagida:1979as}
T.~Yanagida,
``Horizontal gauge symmetry and masses of neutrinos,''
Conf. Proc. C \textbf{7902131}, 95-99 (1979)
KEK-79-18-95.
\bibitem{Yanagida:1979gs}
T.~Yanagida,
``Horizontal Symmetry and Mass of the Top Quark,''
Phys. Rev. D \textbf{20}, 2986 (1979)
 \bibitem{seesaw4}
  R.~N.~Mohapatra and G.~Senjanovic,
  ``Neutrino Mass and Spontaneous Parity Violation,''
  Phys.\ Rev.\ Lett.\  {\bf 44}, 912 (1980).
  
  
  
  
\bibitem{Weinberg:1981wj} 
  S.~Weinberg,
  ``Supersymmetry at Ordinary Energies. 1. Masses and Conservation Laws,''
  Phys.\ Rev.\ D {\bf 26}, 287 (1982).
\bibitem{Sakai:1981pk} 
  N.~Sakai and T.~Yanagida,
  ``Proton Decay in a Class of Supersymmetric Grand Unified Models,''
  Nucl.\ Phys.\ B {\bf 197}, 533 (1982).
  
\bibitem{Goto:1998qg} 
  T.~Goto and T.~Nihei,
  ``Effect of RRRR dimension five operator on the proton decay in the minimal SU(5) SUGRA GUT model,''
  Phys.\ Rev.\ D {\bf 59}, 115009 (1999)
  [hep-ph/9808255].

\bibitem{Super-Kamiokande:2014otb}
K.~Abe \textit{et al.} [Super-Kamiokande],
``Search for proton decay via $p\to\nu K^+$ using 260  kiloton\textperiodcentered{}year data of Super-Kamiokande,''
Phys. Rev. D \textbf{90}, no.7, 072005 (2014)
[arXiv:1408.1195 [hep-ex]].











\bibitem{Matsuda:2000zp} 
  K.~Matsuda, Y.~Koide and T.~Fukuyama,
  ``Can the SO(10) model with two Higgs doublets reproduce the observed fermion masses?,''
  Phys.\ Rev.\ D {\bf 64}, 053015 (2001)
  [hep-ph/0010026].
\bibitem{Matsuda:2001bg} 
  K.~Matsuda, Y.~Koide, T.~Fukuyama and H.~Nishiura,
  ``How far can the SO(10) two Higgs model describe the observed neutrino masses and mixings?,''
  Phys.\ Rev.\ D {\bf 65}, 033008 (2002)
  Erratum: [Phys.\ Rev.\ D {\bf 65}, 079904 (2002)]
  [hep-ph/0108202].
\bibitem{Fukuyama:2002ch} 
  T.~Fukuyama and N.~Okada,
  ``Neutrino oscillation data versus minimal supersymmetric SO(10) model,''
  JHEP {\bf 0211}, 011 (2002)
  [hep-ph/0205066].
\bibitem{Bajc:2002iw} 
  B.~Bajc, G.~Senjanovic and F.~Vissani,
  ``b - tau unification and large atmospheric mixing: A Case for noncanonical seesaw,''
  Phys.\ Rev.\ Lett.\  {\bf 90}, 051802 (2003)
  [hep-ph/0210207].
\bibitem{Goh:2003sy} 
  H.~S.~Goh, R.~N.~Mohapatra and S.~P.~Ng,
  ``Minimal SUSY SO(10), b tau unification and large neutrino mixings,''
  Phys.\ Lett.\ B {\bf 570}, 215 (2003)
  [hep-ph/0303055].
\bibitem{Aulakh:2003kg} 
  C.~S.~Aulakh, B.~Bajc, A.~Melfo, G.~Senjanovic and F.~Vissani,
  ``The Minimal supersymmetric grand unified theory,''
  Phys.\ Lett.\ B {\bf 588}, 196 (2004)
  [hep-ph/0306242].
\bibitem{Goh:2003hf} 
  H.~S.~Goh, R.~N.~Mohapatra and S.~P.~Ng,
  ``Minimal SUSY SO(10) model and predictions for neutrino mixings and leptonic CP violation,''
  Phys.\ Rev.\ D {\bf 68}, 115008 (2003)
  [hep-ph/0308197].
\bibitem{Dutta:2004wv} 
  B.~Dutta, Y.~Mimura and R.~N.~Mohapatra,
  ``CKM CP violation in a minimal SO(10) model for neutrinos and its implications,''
  Phys.\ Rev.\ D {\bf 69}, 115014 (2004)
  [hep-ph/0402113].
\bibitem{Bajc:2004fj} 
  B.~Bajc, G.~Senjanovic and F.~Vissani,
  ``Probing the nature of the seesaw in renormalizable SO(10),''
  Phys.\ Rev.\ D {\bf 70}, 093002 (2004)
  [hep-ph/0402140].
\bibitem{Bertolini:2004eq}
S.~Bertolini, M.~Frigerio and M.~Malinsky,
``Fermion masses in SUSY SO(10) with type II seesaw: A Non-minimal predictive scenario,''
Phys. Rev. D \textbf{70}, 095002 (2004)
[arXiv:hep-ph/0406117 [hep-ph]].
\bibitem{Yang:2004xt}
W.~M.~Yang and Z.~G.~Wang,
``Fermion masses and flavor mixing in a supersymmetric SO(10) model,''
Nucl. Phys. B \textbf{707}, 87-99 (2005)
[arXiv:hep-ph/0406221 [hep-ph]].
\bibitem{Dutta:2004hp}
B.~Dutta, Y.~Mimura and R.~N.~Mohapatra,
``Neutrino masses and mixings in a predictive SO(10) model with CKM CP violation,''
Phys. Lett. B \textbf{603}, 35-45 (2004)
[arXiv:hep-ph/0406262 [hep-ph]].
\bibitem{Dutta:2004zh}
B.~Dutta, Y.~Mimura and R.~N.~Mohapatra,
``Suppressing proton decay in the minimal SO(10) model,''
Phys. Rev. Lett. \textbf{94}, 091804 (2005)
[arXiv:hep-ph/0412105 [hep-ph]].
\bibitem{Babu:2005ia} 
  K.~S.~Babu and C.~Macesanu,
  ``Neutrino masses and mixings in a minimal SO(10) model,''
  Phys.\ Rev.\ D {\bf 72}, 115003 (2005)
  [hep-ph/0505200].
\bibitem{Dutta:2005ni}
B.~Dutta, Y.~Mimura and R.~N.~Mohapatra,
``Neutrino mixing predictions of a minimal SO(10) model with suppressed proton decay,''
Phys. Rev. D \textbf{72}, 075009 (2005)
[arXiv:hep-ph/0507319 [hep-ph]].
\bibitem{Bertolini:2006pe} 
  S.~Bertolini, T.~Schwetz and M.~Malinsky,
  ``Fermion masses and mixings in SO(10) models and the neutrino challenge to SUSY GUTs,''
  Phys.\ Rev.\ D {\bf 73}, 115012 (2006)
  [hep-ph/0605006].
\bibitem{Joshipura:2011nn} 
  A.~S.~Joshipura and K.~M.~Patel,
  ``Fermion Masses in SO(10) Models,''
  Phys.\ Rev.\ D {\bf 83}, 095002 (2011)
  [arXiv:1102.5148 [hep-ph]].
\bibitem{Dueck:2013gca} 
  A.~Dueck and W.~Rodejohann,
  ``Fits to SO(10) Grand Unified Models,''
  JHEP {\bf 1309}, 024 (2013)
  [arXiv:1306.4468 [hep-ph]].
\bibitem{Fukuyama:2015kra} 
  T.~Fukuyama, K.~Ichikawa and Y.~Mimura,
  ``Revisiting fermion mass and mixing fits in the minimal SUSY $SO(10)$ GUT,''
  Phys.\ Rev.\ D {\bf 94}, no. 7, 075018 (2016)
  [arXiv:1508.07078 [hep-ph]].
\bibitem{Fukuyama:2016vgi} 
  T.~Fukuyama, K.~Ichikawa and Y.~Mimura,
  ``Relation between proton decay and PMNS phase in the minimal SUSY $SO(10)$ GUT,''
  Phys.\ Lett.\ B {\bf 764}, 114 (2017)
  [arXiv:1609.08640 [hep-ph]].
\bibitem{Fukuyama:2016mqb} 
  T.~Fukuyama, N.~Okada and H.~M.~Tran,
  ``Sparticle spectroscopy of the minimal SO(10) model,''
  Phys.\ Lett.\ B {\bf 767}, 295 (2017)
  [arXiv:1611.08341 [hep-ph]].
\bibitem{Babu:2016bmy}
K.~S.~Babu, B.~Bajc and S.~Saad,
``Yukawa Sector of Minimal SO(10) Unification,''
JHEP \textbf{02}, 136 (2017)
[arXiv:1612.04329 [hep-ph]].
\bibitem{Babu:2018tfi} 
  K.~S.~Babu, B.~Bajc and S.~Saad,
  ``Resurrecting Minimal Yukawa Sector of SUSY SO(10),''
  JHEP {\bf 1810}, 135 (2018)
  [arXiv:1805.10631 [hep-ph]].
\bibitem{Deppisch:2018flu} 
  T.~Deppisch, S.~Schacht and M.~Spinrath,
  ``Confronting SUSY SO(10) with updated Lattice and Neutrino Data,''
  JHEP {\bf 1901}, 005 (2019)
  [arXiv:1811.02895 [hep-ph]].
\bibitem{Fukuyama:2019zun}
T.~Fukuyama, N.~Okada and H.~M.~Tran,
``Alternative renormalizable $SO$(10) GUTs and data fitting,''
Nucl. Phys. B \textbf{954}, 114992 (2020)
[arXiv:1907.02948 [hep-ph]].
\bibitem{Haba:2020bls}
N.~Haba, Y.~Mimura and T.~Yamada,
``Enhanced $\Gamma(p\to K^0\mu^+)/\Gamma(p\to K^+\bar{\nu}_\mu)$ as a signature of minimal renormalizable SUSY $SO$ (10) GUT,''
PTEP \textbf{2020}, no.9, 093B01 (2020)
[arXiv:2002.11413 [hep-ph]].
\bibitem{Haba:2020ebr}
N.~Haba, Y.~Mimura and T.~Yamada,
``Renormalizable $SO (10)$ grand unified theory with suppressed dimension-5 proton decays,''
PTEP \textbf{2021}, no.2, 023B01 (2021)
[arXiv:2008.05362 [hep-ph]].








\bibitem{Goh:2004fy}
H.~S.~Goh, R.~N.~Mohapatra and S.~Nasri,
``SO(10) symmetry breaking and type II seesaw,''
Phys. Rev. D \textbf{70}, 075022 (2004)
[arXiv:hep-ph/0408139 [hep-ph]].









\bibitem{Nath:2006ut}
P.~Nath and P.~Fileviez Perez,
``Proton stability in grand unified theories, in strings and in branes,''
Phys. Rept. \textbf{441}, 191-317 (2007)
[arXiv:hep-ph/0601023 [hep-ph]].






















 
 
 
\bibitem{Fukuyama:2004xs} 
  T.~Fukuyama, A.~Ilakovac, T.~Kikuchi, S.~Meljanac and N.~Okada,
  ``General formulation for proton decay rate in minimal supersymmetric SO(10) GUT,''
  Eur.\ Phys.\ J.\ C {\bf 42}, 191 (2005)
  [hep-ph/0401213].
\bibitem{Aulakh:2004hm} 
  C.~S.~Aulakh and A.~Girdhar,
  ``SO(10) MSGUT: Spectra, couplings and threshold effects,''
  Nucl.\ Phys.\ B {\bf 711}, 275 (2005)
  [hep-ph/0405074].
\bibitem{Fukuyama:2004ps} 
  T.~Fukuyama, A.~Ilakovac, T.~Kikuchi, S.~Meljanac and N.~Okada,
  ``SO(10) group theory for the unified model building,''
  J.\ Math.\ Phys.\  {\bf 46}, 033505 (2005)
  [hep-ph/0405300].
\bibitem{Fukuyama:2004ti} 
  T.~Fukuyama, A.~Ilakovac, T.~Kikuchi, S.~Meljanac and N.~Okada,
  ``Higgs masses in the minimal SUSY SO(10) GUT,''
  Phys.\ Rev.\ D {\bf 72}, 051701 (2005)
  [hep-ph/0412348].
\bibitem{Bajc:2004xe} 
  B.~Bajc, A.~Melfo, G.~Senjanovic and F.~Vissani,
  ``The Minimal supersymmetric grand unified theory. 1. Symmetry breaking and the particle spectrum,''
  Phys.\ Rev.\ D {\bf 70}, 035007 (2004)
  [hep-ph/0402122].
\bibitem{Bajc:2005qe} 
  B.~Bajc, A.~Melfo, G.~Senjanovic and F.~Vissani,
  ``Fermion mass relations in a supersymmetric SO(10) theory,''
  Phys.\ Lett.\ B {\bf 634}, 272 (2006)
  [hep-ph/0511352].
  
  
 
    
   
   

\bibitem{ParticleDataGroup:2022pth}
R.~L.~Workman \textit{et al.} [Particle Data Group],
``Review of Particle Physics,''
PTEP \textbf{2022}, 083C01 (2022)
   
   
   
   
   
   
\bibitem{FlavourLatticeAveragingGroupFLAG:2021npn}
Y.~Aoki \textit{et al.} [Flavour Lattice Averaging Group (FLAG)],
``FLAG Review 2021,''
Eur. Phys. J. C \textbf{82}, no.10, 869 (2022)
[arXiv:2111.09849 [hep-lat]].
\bibitem{FermilabLattice:2018est}
A.~Bazavov \textit{et al.} [Fermilab Lattice, MILC and TUMQCD],
``Up-, down-, strange-, charm-, and bottom-quark masses from four-flavor lattice QCD,''
Phys. Rev. D \textbf{98}, no.5, 054517 (2018)
[arXiv:1802.04248 [hep-lat]].
\bibitem{Giusti:2017dmp}
D.~Giusti, V.~Lubicz, C.~Tarantino, G.~Martinelli, F.~Sanfilippo, S.~Simula and N.~Tantalo,
``Leading isospin-breaking corrections to pion, kaon and charmed-meson masses with Twisted-Mass fermions,''
Phys. Rev. D \textbf{95}, no.11, 114504 (2017)
[arXiv:1704.06561 [hep-lat]].


\bibitem{EuropeanTwistedMass:2014osg}
N.~Carrasco \textit{et al.} [European Twisted Mass],
``Up, down, strange and charm quark masses with N$_f$ = 2+1+1 twisted mass lattice QCD,''
Nucl. Phys. B \textbf{887}, 19-68 (2014)
[arXiv:1403.4504 [hep-lat]].
\bibitem{Lytle:2018evc}
A.~T.~Lytle \textit{et al.} [HPQCD],
``Determination of quark masses from $\mathbf{n_f=4}$ lattice QCD and the RI-SMOM intermediate scheme,''
Phys. Rev. D \textbf{98}, no.1, 014513 (2018)
[arXiv:1805.06225 [hep-lat]].
\bibitem{Chakraborty:2014aca}
B.~Chakraborty, C.~T.~H.~Davies, B.~Galloway, P.~Knecht, J.~Koponen, G.~C.~Donald, R.~J.~Dowdall, G.~P.~Lepage and C.~McNeile,
``High-precision quark masses and QCD coupling from $n_f=4$ lattice QCD,''
Phys. Rev. D \textbf{91}, no.5, 054508 (2015)
[arXiv:1408.4169 [hep-lat]].


\bibitem{Alexandrou:2014sha}
C.~Alexandrou, V.~Drach, K.~Jansen, C.~Kallidonis and G.~Koutsou,
``Baryon spectrum with $N_f=2+1+1$ twisted mass fermions,''
Phys. Rev. D \textbf{90}, no.7, 074501 (2014)
[arXiv:1406.4310 [hep-lat]].
\bibitem{Hatton:2020qhk}
D.~Hatton \textit{et al.} [HPQCD],
``Charmonium properties from lattice $QCD$+QED : Hyperfine splitting, $J/\psi$ leptonic width, charm quark mass, and $a^c_\mu$,''
Phys. Rev. D \textbf{102}, no.5, 054511 (2020)
[arXiv:2005.01845 [hep-lat]].




\bibitem{Hatton:2021syc}
D.~Hatton, C.~T.~H.~Davies, J.~Koponen, G.~P.~Lepage and A.~T.~Lytle,
``Determination of $\overline{m}_b/\overline{m}_c$ and $\overline{m}_b$ from $n_f=4$ lattice QCD$+$QED,''
Phys. Rev. D \textbf{103}, no.11, 114508 (2021)
[arXiv:2102.09609 [hep-lat]].
\bibitem{Colquhoun:2014ica}
B.~Colquhoun, R.~J.~Dowdall, C.~T.~H.~Davies, K.~Hornbostel and G.~P.~Lepage,
``$\Upsilon$ and $\Upsilon^{\prime}$ Leptonic Widths, $a_{\mu}^b$ and $m_b$ from full lattice QCD,''
Phys. Rev. D \textbf{91}, no.7, 074514 (2015)
[arXiv:1408.5768 [hep-lat]].
\bibitem{ETM:2016nbo}
A.~Bussone \textit{et al.} [ETM],
``Mass of the b quark and B -meson decay constants from N$_f$=2+1+1 twisted-mass lattice QCD,''
Phys. Rev. D \textbf{93}, no.11, 114505 (2016)
[arXiv:1603.04306 [hep-lat]].
\bibitem{Gambino:2017vkx}
P.~Gambino, A.~Melis and S.~Simula,
``Extraction of heavy-quark-expansion parameters from unquenched lattice data on pseudoscalar and vector heavy-light meson masses,''
Phys. Rev. D \textbf{96}, no.1, 014511 (2017)
[arXiv:1704.06105 [hep-lat]].





\bibitem{Sirunyan:2019zvx}
A.~M.~Sirunyan \textit{et al.} [CMS],
``Measurement of $\mathrm{t\bar t}$ normalised multi-differential cross sections in pp collisions at $\sqrt s=13$ TeV, and simultaneous determination of the strong coupling strength, top quark pole mass, and parton distribution functions,''
Eur. Phys. J. C \textbf{80}, no.7, 658 (2020)
[arXiv:1904.05237 [hep-ex]].
\bibitem{ckmfitter}
  J.~Charles {\it et al.} [CKMfitter Group],
  ``CP violation and the CKM matrix: Assessing the impact of the asymmetric $B$ factories,''
  Eur.\ Phys.\ J.\ C {\bf 41}, no. 1, 1 (2005)
  [hep-ph/0406184],
  updated results and plots available at: http://ckmfitter.in2p3.fr
\bibitem{code}
     B.~A.~Kniehl, A.~F.~Pikelner and O.~L.~Veretin,
  ``mr: a C++ library for the matching and running of the Standard Model parameters,''
  Comput.\ Phys.\ Commun.\  {\bf 206}, 84 (2016)
  [arXiv:1601.08143 [hep-ph]]. 
  
  
  
 
\bibitem{Jegerlehner:2001fb} 
  F.~Jegerlehner, M.~Y.~Kalmykov and O.~Veretin,
  ``MS versus pole masses of gauge bosons: Electroweak bosonic two loop corrections,''
  Nucl.\ Phys.\ B {\bf 641}, 285 (2002)
  [hep-ph/0105304];
    F.~Jegerlehner, M.~Y.~Kalmykov and O.~Veretin,
  ``MS-bar versus pole masses of gauge bosons. 2. Two loop electroweak fermion corrections,''
  Nucl.\ Phys.\ B {\bf 658}, 49 (2003)
  [hep-ph/0212319].
\bibitem{Jegerlehner:2003py} 
  F.~Jegerlehner and M.~Y.~Kalmykov,
  ``O(alpha alpha(s)) correction to the pole mass of the t quark within the standard model,''
  Nucl.\ Phys.\ B {\bf 676}, 365 (2004)
  [hep-ph/0308216];
    F.~Jegerlehner and M.~Y.~Kalmykov,
  ``O(alpha alpha(s)) relation between pole- and MS-bar mass of the t quark,''
  Acta Phys.\ Polon.\ B {\bf 34}, 5335 (2003)
  [hep-ph/0310361].
\bibitem{Bezrukov:2012sa} 
  F.~Bezrukov, M.~Y.~Kalmykov, B.~A.~Kniehl and M.~Shaposhnikov,
  ``Higgs Boson Mass and New Physics,''
  JHEP {\bf 1210}, 140 (2012)
  [arXiv:1205.2893 [hep-ph]].
\bibitem{topthreshold}
     P.~Marquard, A.~V.~Smirnov, V.~A.~Smirnov and M.~Steinhauser,
  ``Quark Mass Relations to Four-Loop Order in Perturbative QCD,''
  Phys.\ Rev.\ Lett.\  {\bf 114}, no. 14, 142002 (2015)
  [arXiv:1502.01030 [hep-ph]].
   \bibitem{threshold}
   B.~A.~Kniehl, A.~F.~Pikelner and O.~L.~Veretin,
  ``Two-loop electroweak threshold corrections in the Standard Model,''
  Nucl.\ Phys.\ B {\bf 896}, 19 (2015)
  [arXiv:1503.02138 [hep-ph]]. 
    
   
 
 
   
\bibitem{Machacek:1983tz}
M.~E.~Machacek and M.~T.~Vaughn,
``Two Loop Renormalization Group Equations in a General Quantum Field Theory. 1. Wave Function Renormalization,''
Nucl. Phys. B \textbf{222}, 83-103 (1983)
\bibitem{Machacek:1983fi}
M.~E.~Machacek and M.~T.~Vaughn,
``Two Loop Renormalization Group Equations in a General Quantum Field Theory. 2. Yukawa Couplings,''
Nucl. Phys. B \textbf{236}, 221-232 (1984)
\bibitem{Machacek:1984zw}
M.~E.~Machacek and M.~T.~Vaughn,
``Two Loop Renormalization Group Equations in a General Quantum Field Theory. 3. Scalar Quartic Couplings,''
Nucl. Phys. B \textbf{249}, 70-92 (1985)
   
   
\bibitem{Blazek:1995nv}
T.~Blazek, S.~Raby and S.~Pokorski,
``Finite supersymmetric threshold corrections to CKM matrix elements in the large tan Beta regime,''
Phys. Rev. D \textbf{52}, 4151-4158 (1995)
[arXiv:hep-ph/9504364 [hep-ph]].
   
   
    
    
\bibitem{Esteban:2020cvm}
I.~Esteban, M.~C.~Gonzalez-Garcia, M.~Maltoni, T.~Schwetz and A.~Zhou,
``The fate of hints: updated global analysis of three-flavor neutrino oscillations,''
JHEP \textbf{09}, 178 (2020)
[arXiv:2007.14792 [hep-ph]].
 \bibitem{nufit}
  NuFIT 5.1 (2021), www.nu-fit.org.
    
    
    
    
\bibitem{Borsanyi:2012zv}
S.~Borsanyi, S.~Durr, Z.~Fodor, S.~Krieg, A.~Schafer, E.~E.~Scholz and K.~K.~Szabo,
``SU(2) chiral perturbation theory low-energy constants from 2+1 flavor staggered lattice simulations,''
Phys. Rev. D \textbf{88}, 014513 (2013)
[arXiv:1205.0788 [hep-lat]].


\bibitem{Aoki:2017puj}
Y.~Aoki, T.~Izubuchi, E.~Shintani and A.~Soni,
``Improved lattice computation of proton decay matrix elements,''
Phys. Rev. D \textbf{96}, no.1, 014506 (2017)
[arXiv:1705.01338 [hep-lat]].
 
 
\bibitem{Hisano:1992jj}
J.~Hisano, H.~Murayama and T.~Yanagida,
``Nucleon decay in the minimal supersymmetric SU(5) grand unification,''
Nucl. Phys. B \textbf{402}, 46-84 (1993)
[arXiv:hep-ph/9207279 [hep-ph]].


\bibitem{Super-Kamiokande:2022egr}
R.~Matsumoto \textit{et al.} [Super-Kamiokande],
``Search for proton decay via $p\rightarrow \mu^+K^0$ in 0.37 megaton-years exposure of Super-Kamiokande,''
Phys. Rev. D \textbf{106}, no.7, 072003 (2022)
[arXiv:2208.13188 [hep-ex]].
\bibitem{Super-Kamiokande:2005lev}
K.~Kobayashi \textit{et al.} [Super-Kamiokande],
``Search for nucleon decay via modes favored by supersymmetric grand unification models in Super-Kamiokande-I,''
Phys. Rev. D \textbf{72}, 052007 (2005)
[arXiv:hep-ex/0502026 [hep-ex]].

\bibitem{Hyper-Kamiokande:2018ofw}
K.~Abe \textit{et al.} [Hyper-Kamiokande],
``Hyper-Kamiokande Design Report,''
[arXiv:1805.04163 [physics.ins-det]].

  
\end{thebibliography}
\end{document}